\documentstyle[12pt,graphicx,pslatex,ulem]{article}
\begin{document}

\begin{center}
{\Large {\bf Toroidal modeling of penetration of the resonant magnetic perturbation field
}}

Yueqiang Liu$^1$, A. Kirk$^1$,  and Y. Sun$^2$ 
\end{center}

\vspace{-3mm}
{\it  
$^1$Euratom/CCFE Fusion Association, Culham Science Centre, Abingdon, OX14 3DB, UK\\
$^2$ Institute of Plasma Physics, Chinese Academy of Sciences, 
PO Box 1126, Hefei 230031, China
}

E-mail contact of the main author: yueqiang.liu@ccfe.ac.uk

{\small
{\bf abstract.}
A toroidal, quasi-linear model is proposed to study the penetration
dynamics of the resonant
magnetic perturbation (RMP) field into the plasma. The model couples
the linear, fluid plasma response to a toroidal momentum balance
equation, which includes torques induced by both fluid electromagnetic
force and by
(kinetic) neoclassical
toroidal viscous force. The numerical results for a
test toroidal equilibrium
quantify the effects of various physical parameters on the field penetration and 
on the plasma rotation braking. The neoclassical toroidal viscous torque plays a
dominant role in certain region of the plasma, for the RMP penetration problem
considered in this work.          
}

\section{Introduction}
It is expected that large scale, low frequency type-I edge localized
modes (ELMs) may
not be tolerable for the plasma facing components in ITER, due to the
large heat load \cite{Loarte07}. Extensive experimental results from recent years, on
several existing tokamak devices \cite{Evans06,Liang07,Kirk09,Suttrop11}, have demonstrated that the
externally applied resonant magnetic perturbation (RMP) fields can
significantly affect the behavior of ELMs. It appears that the ELM
mitigation/suppression, and the accompanying density pump-out effect
observed in experiments, require detailed investigations due to complex
physics. 

One particularly important aspect is the RMP field
penetration through the plasma. From the macroscopic point of view,
this is a non-linear dynamic process involving at least two key
effects. One is the plasma response to the applied external field. The
plasma flow has been shown to play a critical role in screening the
RMP field \cite{Waelbroeck03,Heyn08,Becoulet09,Nardon10,LiuPP10}. The other effect
is the rotation braking, due to the plasma response to the external
field. Both fluid (electromagnetic) and kinetic effects can induce
torques damping the plasma flow, in the presence of external
non-axisymmetric fields.  

In this work, we present a fluid-based toroidal, quasi-linear model,
describing the RMP penetration process on the macroscopic scale. The
model couples the plasma response to a toroidal momentum balance
equation, that includes source, sink and diffusion terms. The sink
is provided by the fluid ${\bf j}\times{\bf b}$ torque and the
neoclassical toroidal viscous (NTV) torque. A quasi-linear version
(called MARS-Q) of
the MARS-F code \cite{Liu00} is developed and tested.   
Modeling is carried out
for a test toroidal equilibrium, with mid-plane RMP coils in the
$n=1$ configuration ($n$ is the toroidal mode number).
  
Section 2 describes the quasi-linear model, the numerical
implementation and the benchmark results. Section 3 reports the
modeling results for the test toroidal equilibrium, where a parametric
study is also carried out, in order to clarify the influence of
certain physics parameters on the RMP 
penetration dynamics. Section 4 summarizes the results.
 
\section{Toroidal RMP field penetration model}
The model that we propose here couples the linear plasma response to
the toroidal momentum balance of the plasma. Within the single $n$ 
assumption, the plasma response remains essentially linear. The only non-linear terms come from the
interaction between modes with the same $n$ number, resulting in the
$n=0$ correction to the plasma equilibrium and to the toroidal flow
speed. We neglect the plasma equilibrium correction \cite{Chapman07}, assuming that the
amplitude of the applied RMP field is sufficiently small. The effect
of the RMP field on the
toroidal flow, however, can be significant due to momentum
damping. The damped flow in turn changes the plasma response to the
RMP field. This non-linear coupling is maintained in our model, which
we shall call the quasi-linear RMP penetration model. In what follows,
we describe both components of the model: the plasma response and the
toroidal momentum balance.   

\subsection{Plasma response model}
For the plasma response to the RMP fields, we consider
a resistive, single fluid plasma model, with arbitrary toroidal
flow and flow shear \cite{LiuPP10}. Detailed plasma response
computations have been performed for both MAST and ITER plasmas
\cite{LiuNF11} using this model. 

\begin{eqnarray}
(\frac{\partial}{\partial t}+in\Omega){\bf \xi}&=&{\bf v} + ({\bf\xi}\cdot\nabla\Omega)R\hat\phi, \label{eq:xi}\\
\rho(\frac{\partial}{\partial t}+in\Omega){\bf v}&=&-\nabla p + {\bf j}\times{\bf B} +{\bf
  J}\times{\bf b} - \rho\left[2\Omega\hat{\bf Z}\times{\bf
  v}+({\bf
	 v}\cdot\nabla\Omega)R\hat\phi\right]\nonumber\\
&&-\rho\kappa_{\|}|k_{\|}v_{th,i}| 
  \left[{\bf v}+({\bf\xi}\cdot\nabla){\bf V}_0\right]_{\|}, \label{eq:v}\\ 
(\frac{\partial}{\partial t}+in\Omega){\bf b}&=&\nabla\times({\bf v}\times{\bf B})+({\bf
  b}\cdot\nabla\Omega)R\hat\phi - \nabla\times(\eta{\bf j}), \label{eq:b}\\
(\frac{\partial}{\partial t}+in\Omega)p&=&-{\bf v}\cdot\nabla P-\Gamma
P\nabla\cdot{\bf v}, \label{eq:p}\\ 
{\bf j}&=&\nabla\times {\bf b}, \label{eq:j}
\end{eqnarray}
where $R$ is the plasma major radius, $\hat\phi$ the unit vector along
the geometric toroidal angle $\phi$ of the torus, $\hat{\bf Z}$ the
unit vector in the vertical direction in the poloidal plane. $n$ is the toroidal harmonic
number. The plasma resistivity is denoted by
$\eta$. The variables ${\bf v}, {\bf b}, {\bf j}, p, {\bf \xi}$
represent the perturbed velocity, magnetic field, current, 
pressure, and plasma
displacement, respectively. The equilibrium plasma density, field,
current, and pressure are denoted by $\rho, {\bf B}, {\bf J}, P$,
respectively. $\Gamma=5/3$ is the ratio of specific heats. 

We assume that the plasma equilibrium flow ${\bf V}_0$ has the toroidal component
only, ${\bf V}_0=R\Omega\hat\phi$, with $\Omega$ being the angular
frequency of the toroidal rotation. A parallel sound wave damping term
in added to the momentum equation (\ref{eq:v}), with $\kappa$ being a numerical
coefficient determining the damping ``strength''. $k_{\|}=(n-m/q)/R$
is the parallel wave number, with $m$ being the poloidal harmonic
number and $q$ being the safety factor. $v_{th,i}=\sqrt{2T_i/M_i}$ is
the thermal ion velocity, with $T_i, M_i$ being the thermal ion
temperature and mass, respectively. The parallel component of the perturbed velocity is taken
along the equilibrium field line. The validity of this damping model,
for the RMP field response computations, 
is discussed in Ref. \cite{LiuPP10}.

For the purpose of the RMP response modeling, the
vacuum field equations outside the plasma, the thin resistive wall
equation (when applicable), and the coil equations (Ampere's law) are
solved together with the MHD equations for the plasma.  
The RMP field response modeling requires
solving a linear antenna problem, where the source term is specified
as the current flowing in the magnetic perturbation coils. Since this is a linear
problem, for axi-symmetric equilibria, we only need to consider a
single toroidal mode number $n$ at one time. Therefore, the source
current is assumed to have an $\exp(in\phi)$ dependence along the toroidal
angle $\phi$. 

\subsection{Toroidal momentum balance model}
The toroidal momentum equation is derived from the
force balance equation
\begin{eqnarray}
\rho\frac{\partial{\bf V}}{\partial t} = {\bf J}\times{\bf B} - \nabla
P - \nabla\cdot{\bf \pi} + {\bf S}. \label{eq:S}
\end{eqnarray}
where ${\bf V}$ is the plasma flow velocity, ${\bf\pi}$ the viscous
tensor, and ${\bf S}$ denoting the source term for the force. 

Following Ref. \cite{ShaingNF10}, the flux surface averaged toroidal
moment $L=\rho<R^2>\Omega$ satisfies  
\begin{eqnarray}
\frac{\partial L}{\partial t}  = D(L) + T_{NTV}(\omega_E)
+ T_{j\times b} + T_{\rm source},  \label{eq:mom}
\end{eqnarray}
where $\omega_E$ is the toroidal $E\times B$ drift frequency. The
toroidal torque, due to the generalized viscous force $\nabla\cdot{\bf
  \pi}$, is split into three terms: the momentum
diffusion and pinch term $D$, the toroidal component of the
neoclassical toroidal viscosity (NTV) torque
$T_{NTV}$ and the fluid electromagnetic torque $T_{j\times b}$. The
first term can be written as  \cite{SunPPCF10}
\begin{eqnarray*}
D = \frac Gs\frac{\partial}{\partial
  s}\frac sG \left[\chi_M<|\nabla s|^2>\frac{\partial
	 L}{\partial s} + V_{\rm pinch}<|\nabla
  s|>L\right], \quad G\equiv F<1/R^2>,
\end{eqnarray*}
 where $s$ labels the radial coordinate, $F$ is the equilibrium poloidal current flux function, $\chi_M$
 the (anomalous) toroidal momentum diffusion coefficient, and $V_{\rm
	pinch}$ the pinch velocity. 

The torque $T_{\rm source}$ from Eq. (\ref{eq:mom}) comes from
the source force term ${\bf S}$ in Eq. (\ref{eq:S}), denoting, for
instance, the momentum input due to the neutral beam injection.

The surface averaged, toroidal electromagnetic ${\bf j}\times{\bf b}$ torque
density is computed as 
\begin{eqnarray*}
T_{j\times b}=\oint R{\bf j}\times{\bf b}\cdot\hat\phi dS/\oint dS,
\end{eqnarray*}
where $R$ is the major radius, ${\bf j}$ and ${\bf b}$ are the (total)
perturbed plasma current and magnetic field, respectively. $S$ denotes
the flux surface. It should be pointed out that the total toroidal torque,
acting on the plasma column,
can be either computed by integrating the torque density defined
in the above equation across the whole plasma minor radius, or by
direct evaluation of a surface integral, at an arbitrary surface
in the vacuum region between the plasma boundary and the first
conducting structure. The integrand of the surface integral is the
product of the perturbed radial and toroidal field components
only \cite{Pustovitov07}. These two equivalent methods provide an
internal check of the numerical implementation for the ${\bf
  j}\times{\bf b}$ torque density calculation. This internal check
has been successfully performed in the MARS-Q code.     

The NTV torque is computed here using formulas from Ref. \cite{ShaingNF10},
where various regimes (the so-called {$\nu-\sqrt{\nu}$ and $1/\nu$ regimes, as
well as the superbanana and superbanana plateau regimes}) are smoothly connected. We point out that
these formulas do
not treat the exact pitch angle scattering operator, nor the particle
resonance effects {associated with the} bounce frequency
\cite{ParkPRL09}. Despite this, the approximate formulas from
Ref. \cite{ShaingNF10} are {reasonably well verified by numerical results} 
\cite{SunPRL10}. {Comparison of this NTV theory with experimental data
in JET \cite{SunNF12} and DIII-D \cite{LiuPPCF12} shows better than the order of
magnitude agreement, as long as the plasma response is properly taken into
account in computing the torque.}

If we assume that a momentum balance has been achieved before applying
the RMP field, with $(\Omega_0, L_0, \omega_E^0)$ satisfying 
\begin{eqnarray*}
D(L_0) + T_{source} = 0.
\end{eqnarray*}

After applying the RMP field (without changing other equilibrium conditions), we define 
\begin{eqnarray*}
\Omega(t) = \Omega_0 + \Delta\Omega(t), \qquad L(t) = L_0 + \Delta
L(t), \qquad \omega_E = \omega_E^0 + \Delta\omega_E = \omega_E^0 +
\Delta\Omega, 
\end{eqnarray*}
and obtain the following momentum balance equation in the presence
of RMPs
\begin{eqnarray}
\frac{\partial \Delta L}{\partial t}  = D(\Delta L) +
T_{NTV}(\omega_E^0+\Delta\Omega) + T_{j\times b}, \label{eq:momf}
\end{eqnarray}
which is solved in MARS-Q, together with the linear MHD equations
describing the plasma response to the RMP field. In the presence of
the diffusion operator, equation
(\ref{eq:momf}) requires two boundary conditions, at the plasma center
and edge, respectively. 
We use a Neumann type of
boundary condition $\partial\Delta L/\partial s=0$ at the plasma
center. At the plasma edge, we assume a homogeneous 
Dirichlet boundary condition for $\Delta
L$. For tokamak plasmas, this is a reasonable approximation of the
more generic Robin boundary condition, as demonstrated in
Ref. \cite{FitzpatrickNF93}, by considering a thin scrape-off layer 
surrounding the plasma.

It is now the proper time to discuss the validity of the above proposed
quasi-linear model for the RMP field penetration
computations. Obviously this is essentially a single fluid model, especially
for the plasma response part. Inclusion of two fluid effects
\cite{Yu09,Nardon10,Ferraro12}, as well as kinetic effects \cite{Heyn08,Park10} into the plasma
response, remains our future work. In this work, we try to understand
the MHD aspects of the RMP field penetration, by including the NTV
torque into the momentum balance, and by considering a full toroidal
geometry. 

The other question is the validity of the model in terms of the time
scale. Both experimental evidence and modeling results
\cite{Park10}, including those to be shown in this work, seem to
suggest that the RMP penetration occurs at the time scale of several
milliseconds, which is much slower than the Alfv\'enic time, but faster
than the plasma resistive diffusion time. Therefore, at this time scale, we argue that the
linear resistive response of the plasma, without inclusion of the
finite island width effect, is appropriate. This is essentially the
thin-island approximation, which is invalid for fully reconnected,
large magnetic islands. Such islands form after the full penetration
of the RMP field into the plasma. 

On the other hand, we do
not need to model the details of the Alfv\'en wave dynamics, which can
be avoided by choosing a fully implicit time-stepping scheme for the
full MHD equations. This time-stepping scheme is described in the
following Subsection.   

\subsection{Time-stepping scheme for solving quasi-linear equations}\label{sec:adap}
The coupled MHD-momentum balance equations can be symbolically written
as 
\begin{eqnarray*}
B\frac{\partial X}{\partial t} &=& A_1X + YA_2 X + X_0, \\
C\frac{\partial Y}{\partial t} &=& DY + T(Y)X^2,
\end{eqnarray*}
where the first equation is the full linearized MHD equation; with $X$
denoting the full set of the existing MARS-F solution 
variables; $Y\equiv \Delta\Omega$ being the modification of the
toroidal rotation frequency due to various torques, $A_1$ denoting the
MHD operator, that also
contains the initial rotation $\Omega_0$; $X_0$ denoting the
source term, i.e. the RMP current. 

The second equation above is the momentum balance equation for
$Y$. The first term from the right hand side denotes the linear momentum diffusion-pinch
term. The second term from the right hand side denotes all the torque terms, with the
coefficient $T$ being generally a non-linear function of $Y$. The
quadratic dependence of torques on the MHD perturbation variable $X$
reflects the fact that the product of two $n\neq 0$ perturbations (the
plasma current and the magnetic field) results in the $n=0$ torque. 

MARS-Q uses the following time stepping scheme based on a staggered
grid in time 
\begin{eqnarray*}
&&B\frac{X^{k+1}-X^k}{\Delta t} = (1-\alpha_2)A_1X^k + \alpha_2A_1X^{k+1} +
(1-\alpha_3)Y^{k+1/2}A_2X^k + \alpha_3Y^{k+1/2}A_2X^{k+1} + X_0, \\
&&C\frac{Y^{k+1/2}-Y^{k-1/2}}{\Delta t}= (1-\alpha_6)DY^{k-1/2} + \alpha_6DY^{k+1/2} + T(Y^{k-1/2})\left(X^{k+1}\right)^2.
\end{eqnarray*}
where $\alpha_i, i=1,\cdots 6$, are coefficients determining the
numerical scheme of time stepping. 
We shall consider the RMP field penetration process (ms time scale) that is much
faster than the Alfv\'en time $\tau_A\equiv R_0\sqrt{\mu_0\rho_0}/B_0$
($R_0,\rho_0,B_0$ are the major radius, the plasma density, and the
toroidal magnetic field at the plasma center, respectively), which is
normally in the $\mu$s scale. This allows us to neglect the
detailed dynamics of fast Alfv\'en waves, that can be achieved by
choosing a fully implicit time-stepping scheme for the MHD operators,
i.e. $\alpha_2=\alpha_3=1$, and by choosing the time step $\Delta t$
larger than 1. Our numerical computations for the test toroidal
equilibrium show that $\Delta t$ can be as large as $10\tau_A$,
without compromising numerical accuracy for the time trace, as will be
shown later.  Normally for time-stepping the momentum equation, we
also choose the fully implicit scheme $\alpha_6=1$ for the linear
operators. 

We also designed a simple adaptive time-stepping scheme for
solving the fully coupled equations, in which the time step 
depends on the iteration $\Delta t=\Delta t_k$. During the time-stepping, the
code computes a quantity $\delta$, characterizing the relative change of the
solution (e.g. the $n\neq0$ plasma response field and displacement)
between two consecutive time steps. If $\delta$ is larger than a
prescribed parameter $\delta_{\max}$, the next time step is reduced by
a factor $\alpha_7<1$, i.e. $\Delta t_{k+1}=\alpha_7\Delta t_k$.  If
$\delta$ is smaller than a 
prescribed parameter $\delta_{\min}$, the next time step is increased by
the factor $1/\alpha_7$. For the modeling results shown in Sections
\ref{sec:tor}, where the time adaptivity is applied, we
choose $\delta_{\max}=10$\%, $\delta_{\min}=2$\%, and $\alpha_7=0.8$.    

\subsection{Benchmarking the momentum solver}
The final momentum equation (\ref{eq:momf}) is solved using a finite
element method (FEM) along the radial grid. For simplicity, we assume homogeneous Neumann boundary
conditions for $\Delta L$ at both the plasma center and edge in this
analytic benchmark. [We note, though, that for physical problems to be
solved in Section 3, we assume the Dirichlet boundary condition at the
plasma edge.] With a given source term $T$ which does not depend on
time $t$ and the solution $y$, Eq. (\ref{eq:momf}) has a
general form of
\begin{eqnarray}
c\frac{\partial y}{\partial t} = \frac1a\frac{\partial}{\partial s}
a\left(b\frac{\partial y}{\partial s} + dy\right) + T, \label{eq:anaeq}
\end{eqnarray}
which allows an analytic steady state solution (which generally exists
except some trivial cases) 
\begin{eqnarray*}
y(s)|_{t\to\infty}&=&\int_0^s\frac{e^{\alpha(t)-\alpha(s)}}{ab}\left[a_0d_0y_0-\int_0^taTdu\right]dt
  + y_0e^{-\alpha(s)},\\
y_0&=&\left[a_1d_1a_0d_0\int_0^1\frac{e^{\alpha(t)-\alpha(1)}}{ab}dt +
  a_1d_1e^{-\alpha(1)} - a_0d_0\right]^{-1}\times \\
&& \left[
  a_1d_1\int_0^1\frac{e^{\alpha(t)-\alpha(1)}}{ab}dt\int_0^taTdu - \int_0^1aTdt\right], \\
  \alpha(s)&\equiv& \int_0^s\frac dbdt,
\end{eqnarray*}
This analytic solution is used to test the FEM momentum solver in
MARS-Q. A special case is considered, with      
\begin{eqnarray*}
a(s) = a_0e^{\beta s}, \quad b=b_0, \quad d=d_0, \quad
\frac{d_0}{b_0}=\alpha, \quad T=T_0e^{\gamma s},
\end{eqnarray*}
and the steady state solution
\begin{eqnarray}
y(s) &=& y_0\frac{\alpha e^{-\beta s}-\beta e^{-\alpha s}}{\alpha-\beta}
- \frac{T_0}{d_0}\frac{\alpha}{\beta+\gamma}\left(\frac{e^{\gamma
	 s}-e^{-\alpha s}}{\alpha+\gamma} - \frac{e^{-\beta s}-e^{-\alpha
	 s}}{\alpha-\beta}\right), \label{eq:anay}\\
y_0&=&
\frac{T_0}{d_0}\frac{\alpha-\beta}{\beta(\beta+\gamma)}\left(\frac{\alpha}{\alpha+\gamma}
- \frac{\alpha}{\alpha-\beta} +
\frac{\gamma}{\alpha+\gamma}\frac{e^{\gamma}-e^{-\beta}}{e^{-\alpha}-e^{-\beta}}\right). 
\end{eqnarray}

\begin{figure}
\begin{center}
\includegraphics[width=6cm]{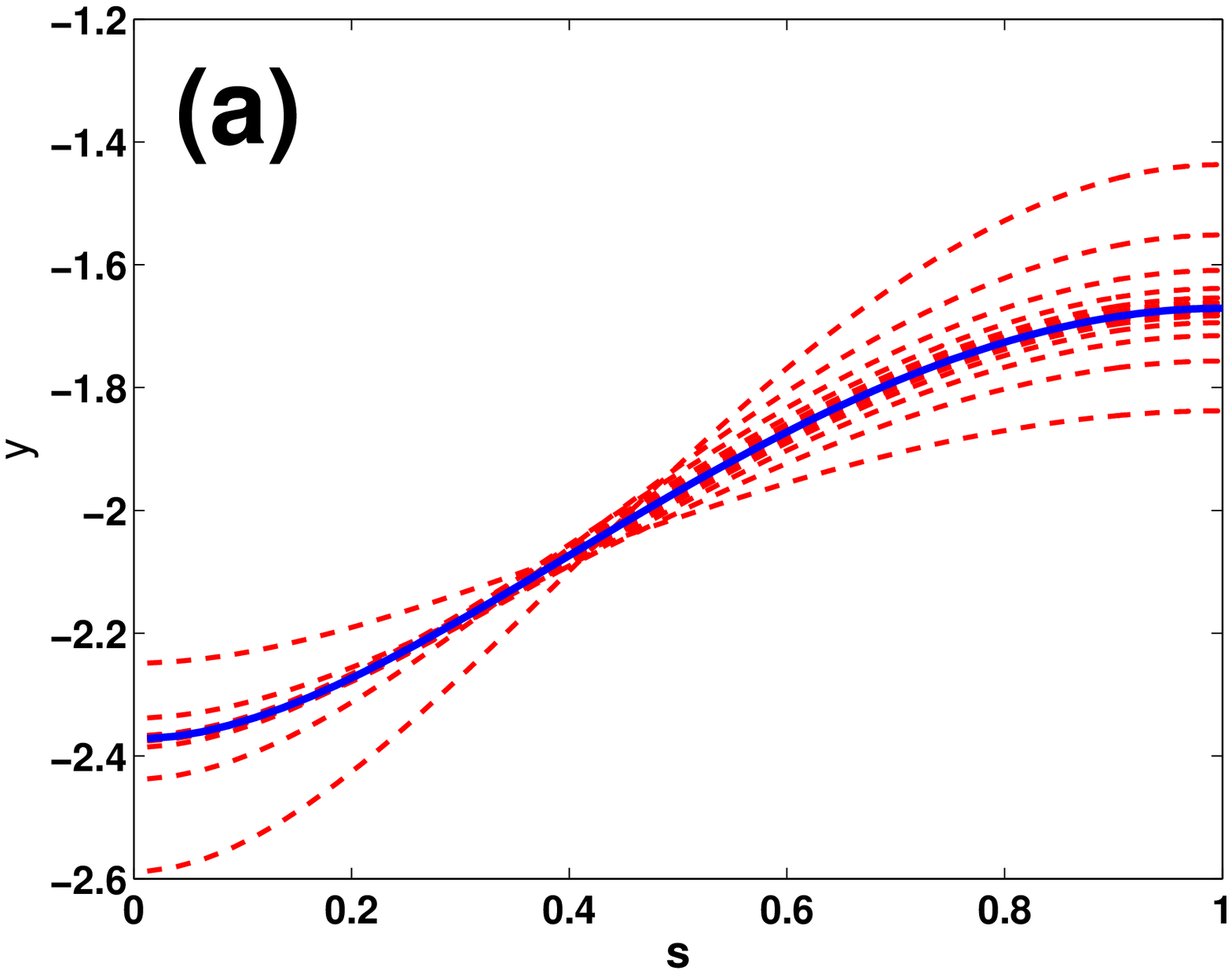}
\includegraphics[width=6cm]{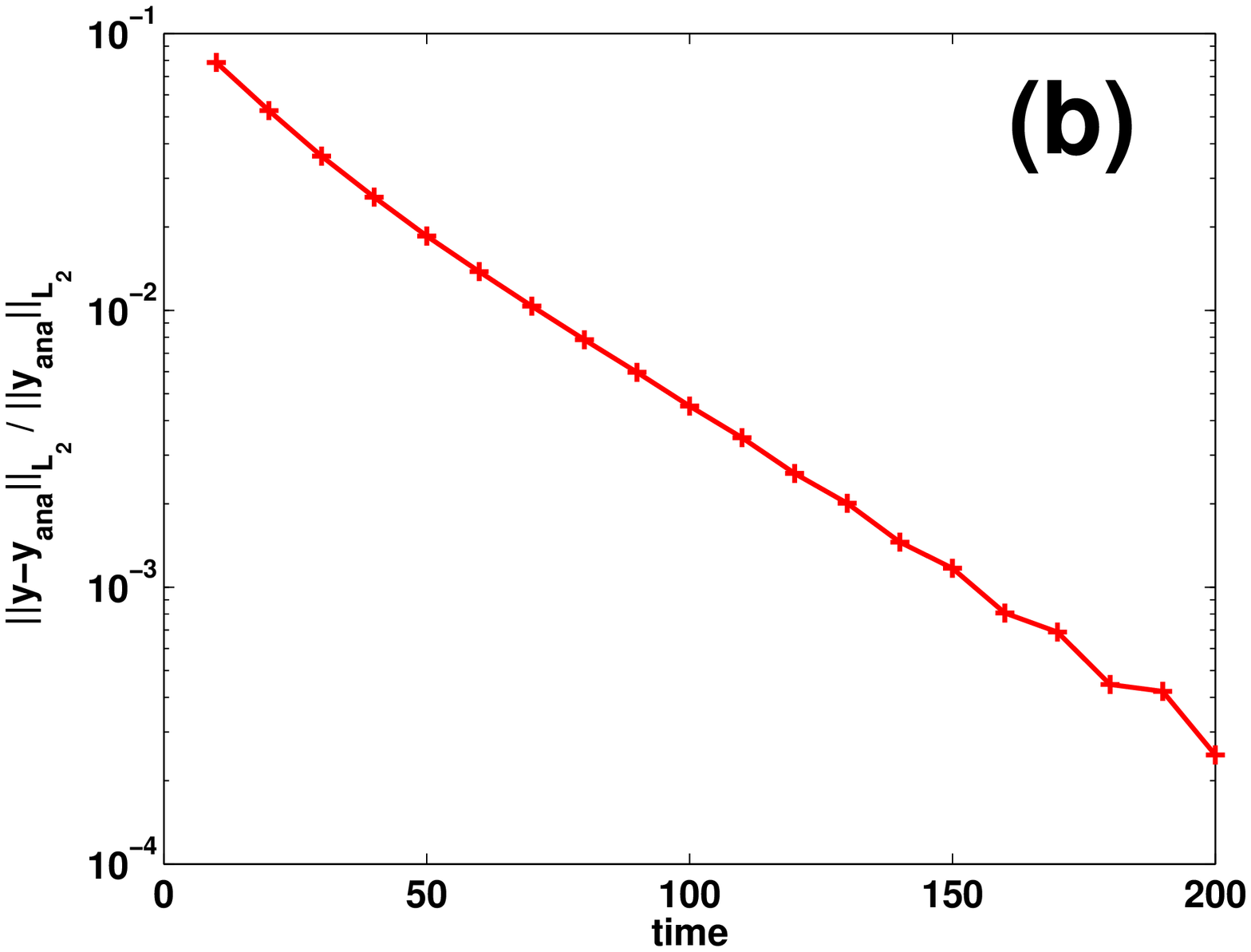}
\caption{ Test of the MARS-Q momentum solver against analytic
  solution (\ref{eq:anay}), for a case with $a_0=2, d_0=3, T_0=3.2,
  \alpha=1.5, \beta=2.3, \gamma=1.7$. Shown are (a) the convergence of
  the numerical profiles (dashed) to the analytic profile (solid), and
  (b) the convergence of the relative error of the solution, in $L_2$
  norm, to the steady state analytic solution. The convergence of the
  radial profiles, shown in (a), comes from both sides of the dashed line, in
  an oscillating manner. The time step is chosen $\Delta
  t=10$, with the implicity parameter $\alpha_6$=0.6.} 
\label{fig:anatest}
\end{center}
\end{figure}

Figure \ref{fig:anatest} shows an example of the MARS-Q computed time
evolution of Eq. (\ref{eq:anaeq}), with the coefficient $c=1$, the
time step $\Delta t=10$, and the implicity parameter $\alpha_6$=0.6.
The numerical
solution converges to the analytic steady state solution. The
convergence speed depends on the choice of parameter $\alpha_6$. At a
given $\Delta t$, larger $\alpha_6$ (i.e. more ``implicit'' scheme)
usually gives faster convergence. Note that, since
Eq. (\ref{eq:anaeq}) represents a pure mathematical model, no specific physical
units are associated with all the quantities here.   

\section{Numerical results for a test toroidal equilibrium} \label{sec:tor}
\subsection{Equilibrium and RMP field configuration}
The MARS-Q code allows quasi-linear simulations of the RMP
field penetration dynamics and the plasma toroidal momentum damping,
by coupling the $n\neq0$ perturbed, full MHD equations with the $n=0$ toroidal
momentum balance equation. The modeling is performed for full
toroidal geometry. The NTV torque is included into the momentum
balance equation. Only toroidal plasma flow is considered. These are
the major difference from a previous 
work \cite{Nardon10}, based on a four-field reduced MHD model, and 
cylindrical geometry.

\begin{figure}
\begin{center}
\includegraphics[width=12cm]{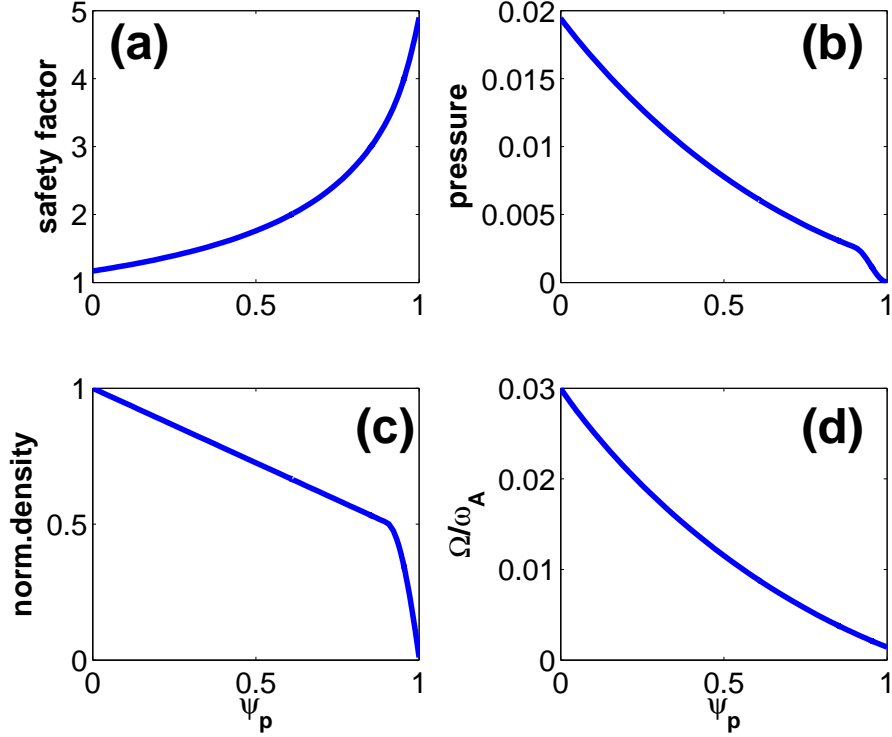}
\caption{The radial profiles of the safety factor $q$, the equilibrium
  pressure (normalized by $B_0^2/\mu_0$), the normalized plasma density (to unity at the magnetic
  axis), and the plasma toroidal rotation frequency $\Omega$, for a
  test toroidal equilibrium.}  
\label{fig:eqprof}
\end{center}
\end{figure}

We consider an analytic specification of the radial profiles for a
toroidal equilibrium \cite{LiuPP10}, in which the
equilibrium current and pressure profiles, as well as the plasma
boundary shape is specified analytically. The key radial profiles are
shown in Fig. \ref{fig:eqprof}. 
The plasma major radius of $R_0=3$m, the vacuum toroidal magnetic
field $B_0$=1.5Tesla, and the aspect ratio $R_0/a=3$. The plasma boundary
has an elongation $\kappa=1.6$ and triangularity $\delta=0.3$. The
equilibrium current and pressure are chosen to have $q_0=1.17,
q_{95}=3.94, q_a=4.90$, and the normalized pressure
$\beta_N=1.56$. This plasma is far
below the no-wall limit for the $n=1$ ideal external kink
instability. The total plasma current is 1.37MA. 

For test computations, we consider the RMP field produced by a set of
4 coils located at $(R,Z)=(4.98,1)$m and $(4.98,-1)$m. These coils are
uniformly distributed along the
toroidal angle, each covering 90$^{\rm o}$ toroidal angle. The coils are
outside a resistive wall located at the minor radius of $1.23a$,
resembling the error field correction coils (EFCC) in JET. 
The polarity of the coil currents are arranged to produce a
predominantly $n=1$ RMP field. 

\subsection{Numerical results for the base case}
In order to investigate the effect of various physical and numerical
parameters on the dynamics of the field penetration and the rotation
damping, we first define a base case as follows. 
We consider a
resistive plasma with the magnetic Lundquist number $S=10^8$ at the
magnetic axis. The radial profile of the plasma resistivity scales as
$T_e^{-3/2}$, where $T_e$ is the equilibrium thermal electron
temperature. This leads to the $S$-value of about $10^6$ near the
plasma edge. 
We choose an amplitude of the anomalous toroidal momentum diffusion
coefficient $\chi_M^0=3\times10^{-7}R_0v_A\simeq5$m$^2$/s, similar to
the value in a typical JET plasma \cite{Baranov09}. The radial profile
of the momentum diffusion coefficient varies between two somewhat extreme
examples. In the first example, which is used for the base case,
$\chi_M(\psi_p)=\chi_M^0\psi_p^{-/2}$. This gives a larger momentum
diffusion in the plasma core than in the edge. The other example, to
be used later in this work, is 
$\chi_M(\psi_p)=\chi_M^0[T_e(\psi_p)/T_e(0)]^{-3/2}$, which gives a
larger momentum diffusion in the edge than in the core.
The pinch velocity is neglected in this work.  For the base case, both
the $j\times b$ and NTV torques are included in the momentum
equation.  Finally, 
we assume that each of the RMP coils carries a
20kAt current. 

\begin{figure}
\begin{center}
\includegraphics[width=6.5cm]{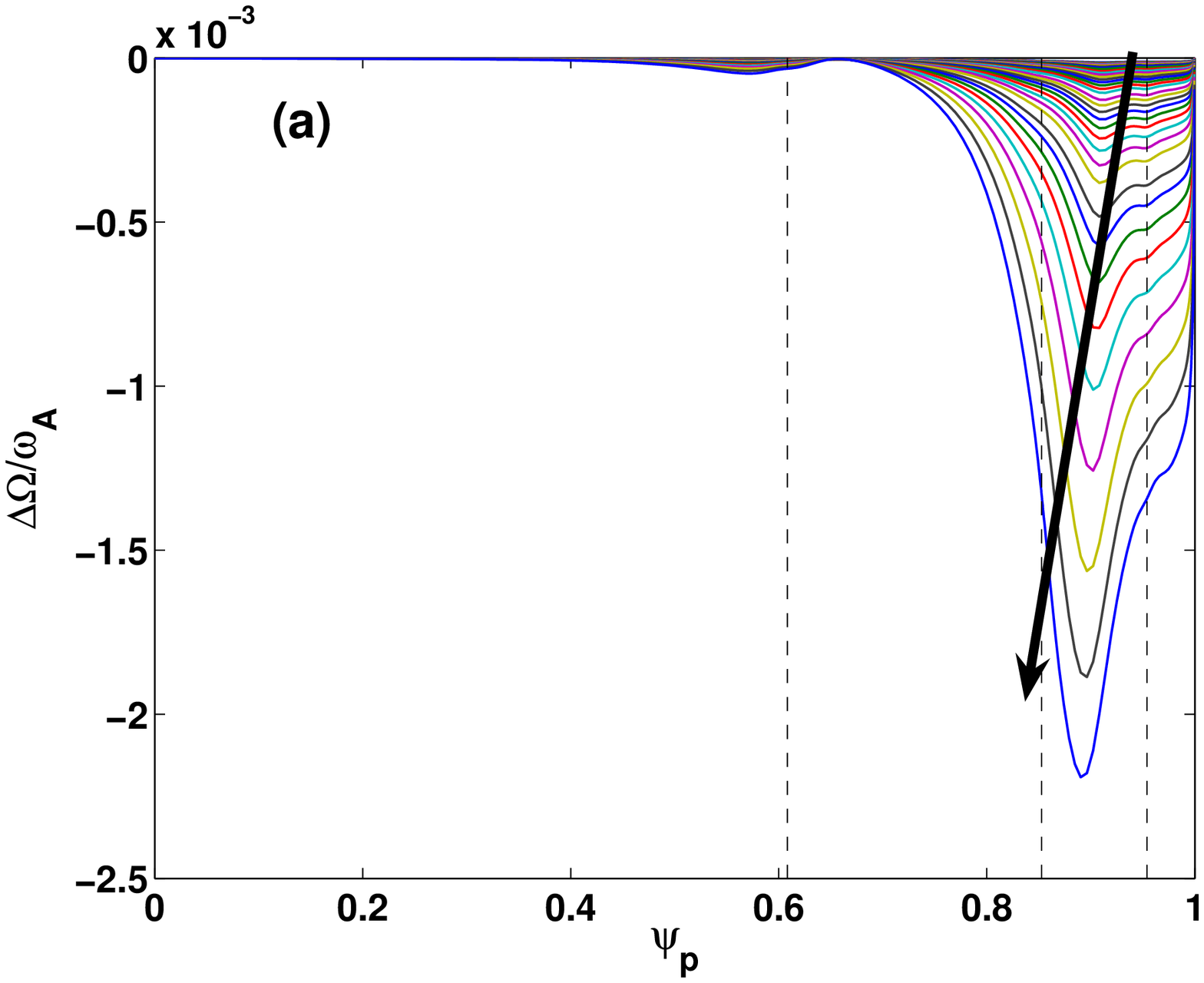}
\includegraphics[width=6.5cm]{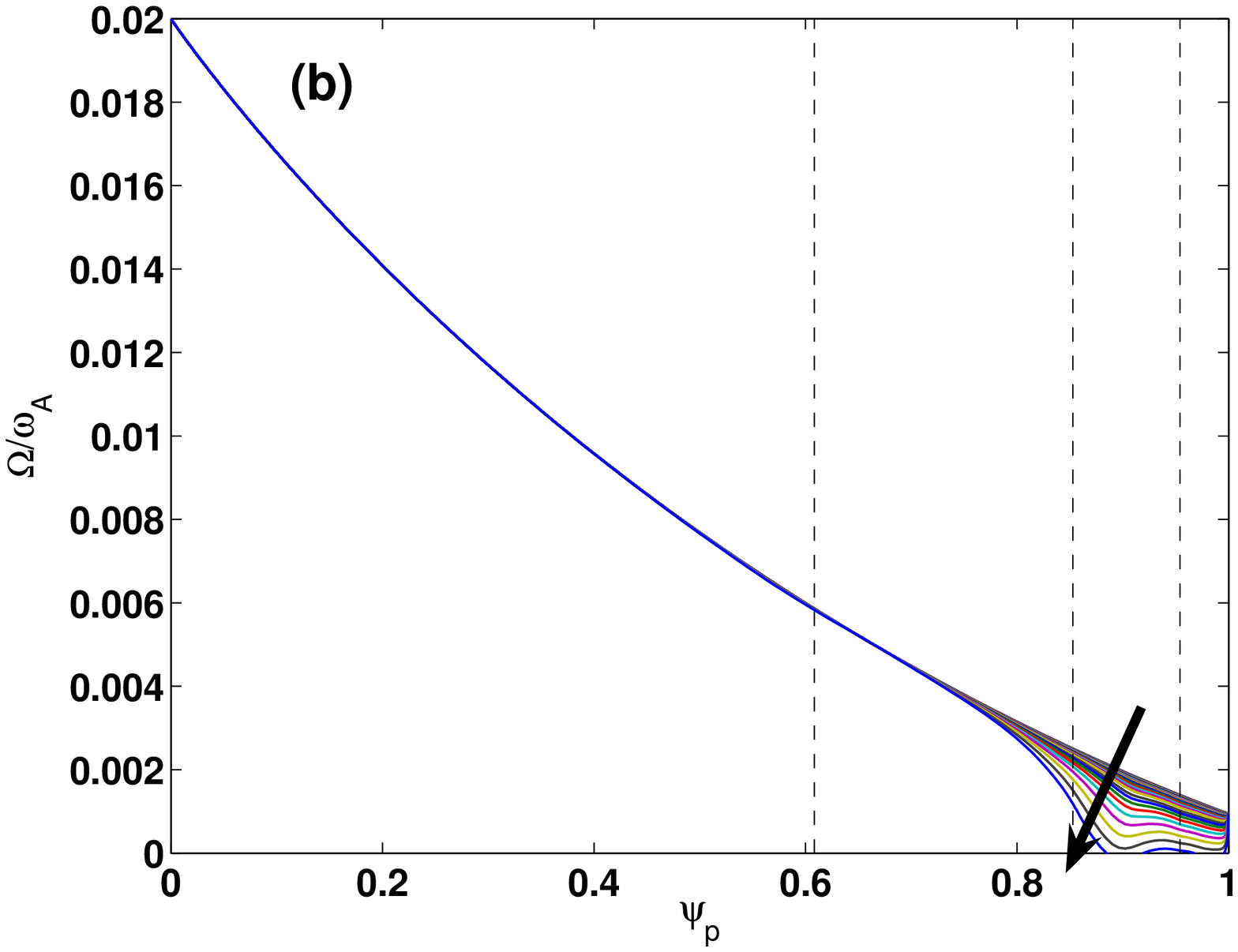}
\caption{Evolution of the simulated radial profiles of (a)
  $\Delta\Omega(\psi_p,t)\equiv\Omega(\psi_p,t)-\Omega(\psi_p,t=0)$ and (b) 
  $\Omega(\psi_p,t)$ for the base case, where $\Omega$ is the toroidal 
  rotation frequency, $\psi_p$ is the normalized equilibrium poloidal
  flux, and $t$ is the time.  Shown are only profiles with a
  time span of 0.1ms, and after 10ms of simulation. The arrow
  indicates the time flow. The vertical dashed lines indicate radial
  locations of the $q=2,3,4$ rational surfaces, respectively.} 
\label{fig:basewp}
\end{center}
\end{figure}

The direct consequence of the non-linear interaction between the plasma
response (to the RMP fields) and the plasma flow is the flow damping,
which is the primary effect that we report in this work. Figure
\ref{fig:basewp} shows the evolution of the radial profile of the
toroidal rotation frequency during this non-linear interaction, for
the plasma and coil configurations as described for the base case. We
obtain generally a full braking of the plasma flow near the edge region (beyond
the $q=3$ surface). A full penetration of the RMP field, into the
plasma edge region, is expected as
the rotation vanishes in that region. At full penetration, large
magnetic islands form, which in turn invalidates the thin-island
assumption used in the MARS-Q model. Therefore, generally speaking,
our numerical results 
are valid only for the time interval before the full braking of the
toroidal flow.  
We also note that, at the moment of the full rotation braking beyond
the $q=3$ surface, the core plasma rotation is still well maintained. 

For this base case, as well as for other cases presented in this
work, further time stepping does not yield a steady state
solution. One possible reason is the violation of the quasi-linear
assumption in the model, as discussed above. The other possibility is
the developement of (non-linear) MHD instabilities near the plasma edge region,
where both the rotation and rotation shear exhibit rapid
changes. Allowing even further time evolution, the simulation produces
numerically incorrect results. Therefore, for cases where no steady
state solutions are reached, the physically meaningful solution is the
time evolution before the full braking of the edge rotation of the
plasma. This is also the physically interesting solution since it
represents the dynamic process of the RMP field penetration. We
mention that for certain plasmas, steady state solutions can be
obtained by the MARS-Q quasi-linear model. Examaples can be found from
Ref. \cite{LiuPPCF12}.
      
\begin{figure}
\begin{center}
\includegraphics[width=6.5cm]{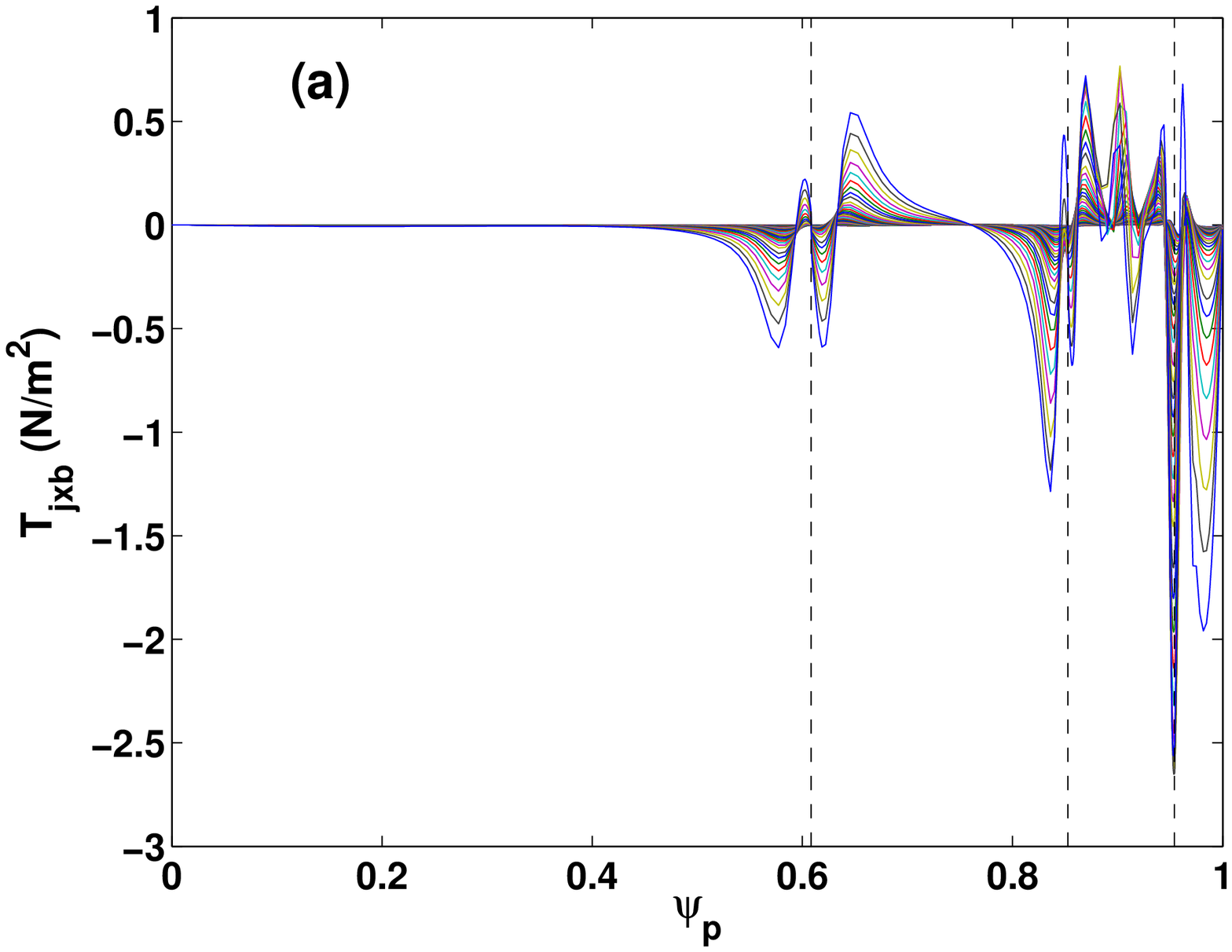}
\includegraphics[width=6.5cm]{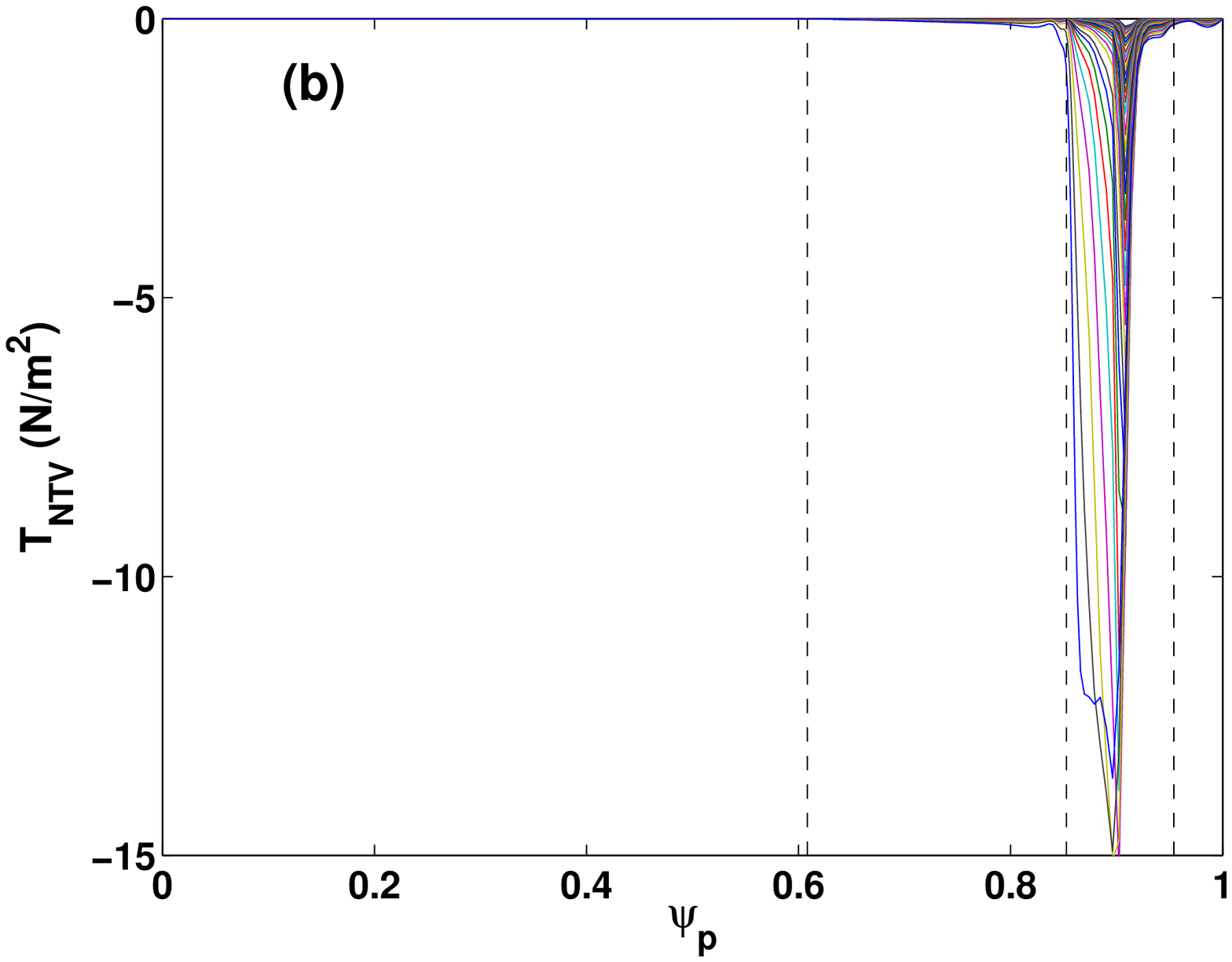}
\caption{Evolution of the simulated radial profiles of (a) the
  electromagnetic torque density and (b) the NTV torque density
  for the base case.  Shown are only profiles with a
  time span of 0.1ms, and after 10ms of simulation. The vertical
  dashed lines indicate radial 
  locations of the $q=2,3,4$ rational surfaces, respectively.} 
\label{fig:basetorq}
\end{center}
\end{figure}

\begin{figure}
\begin{center}
\includegraphics[width=8cm]{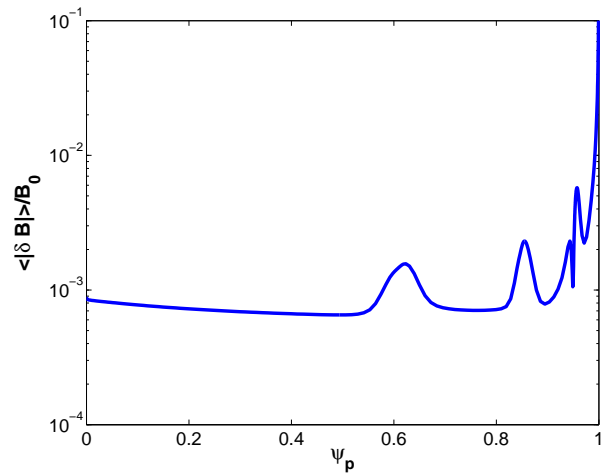}
\caption{{The radial profile of the flux surface averaged magnetic
	 field strength including the plasma response, at the initial toroidal flow
	 speed.}}  
\label{fig:basedeltab}
\end{center}
\end{figure}

The observed rotation braking is caused by the electromagnetic and the
NTV torques, whose radial profile evolution is shown in
Fig. \ref{fig:basetorq}. Note that the ${\bf j}\times{\bf b}$ torque,
though mainly occurring near rational surfaces, is nevertheless
distributed along the minor radius, with non-trivial profiles. This is
partially due to the continuum resonance induced splitting effect as
discussed in \cite{LiuPP12b}. The NTV torque, for the case considered
here, is mainly localized between the $q=3$ and 4 rational
surfaces. This is in fact the major factor braking the plasma rotation
between the $q=3$ and 4 rational surfaces, as will be shown later
(Fig. \ref{fig:emwp}). However, we point out that this type of the NTV torque
distribution, observed in most of the computations for the plasma
studied in this work, should not be regarded as a ubiquitous feature
valid for
any plasma equilibria. The NTV torque is generally a rather non-linear
function of the plasma ${\bf E}\times{\bf B}$ flow. In addition, the
torque distribution also depends on the radial profile of the plasma
collisionality, the drift kinetic resonance between the plasma
response and plasma thermal particles, and finally on the spacial
distribution of the
perturbed 3D field amplitude $|\delta{\bf B}|$. All these factors can potentially
affect the eventual radial profile of the NTV torque density. {
  Figure \ref{fig:basedeltab} shows one example of the flux surface
  averaged $|\delta{\bf B}|$, normalized by the vacuum toroidal field
  at the magnetic axis, computed for the plasma response with the
  initial flow speed. The field amplitude predominantly comes from the
  Lagrangian variation (i.e. the field variation on the distorted flux
  surface). The computed field strength is of order of $10^{-3}$ of
  the vacuum field in
  the major part of the plasma column, but is larger near the plasma
  boundary, due to the larger plasma displacement towards the
  edge.} More 
toroidal examples (and discussions of the above factors) are found
in Ref. \cite{LiuPPCF12}. For the case considered here, we note that
the amplitude of the NTV torque density is roughly about 5 times
larger than that of the electromagnetic torque.       

\begin{figure}
\begin{center}
\includegraphics[width=6.5cm]{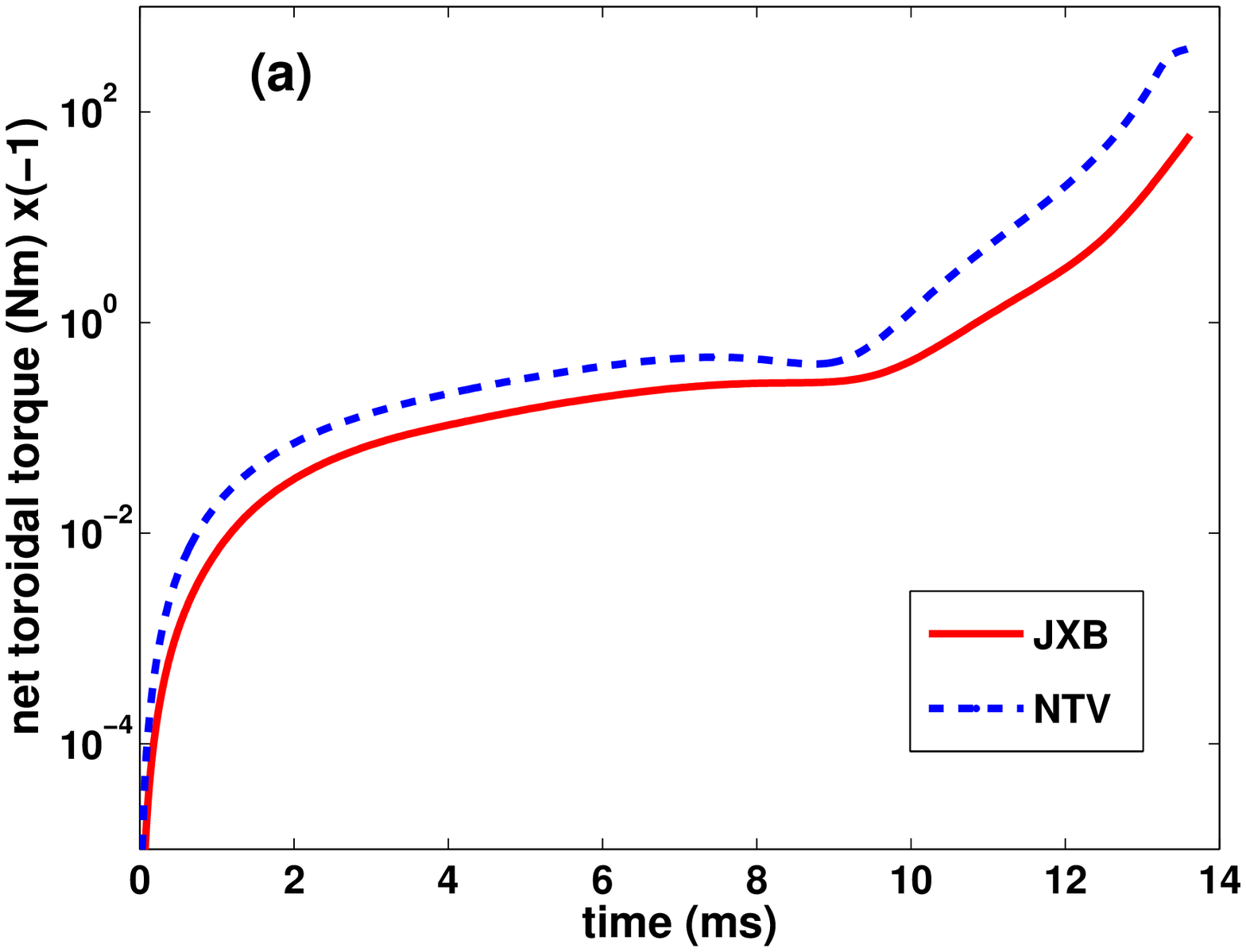}
\includegraphics[width=6.5cm]{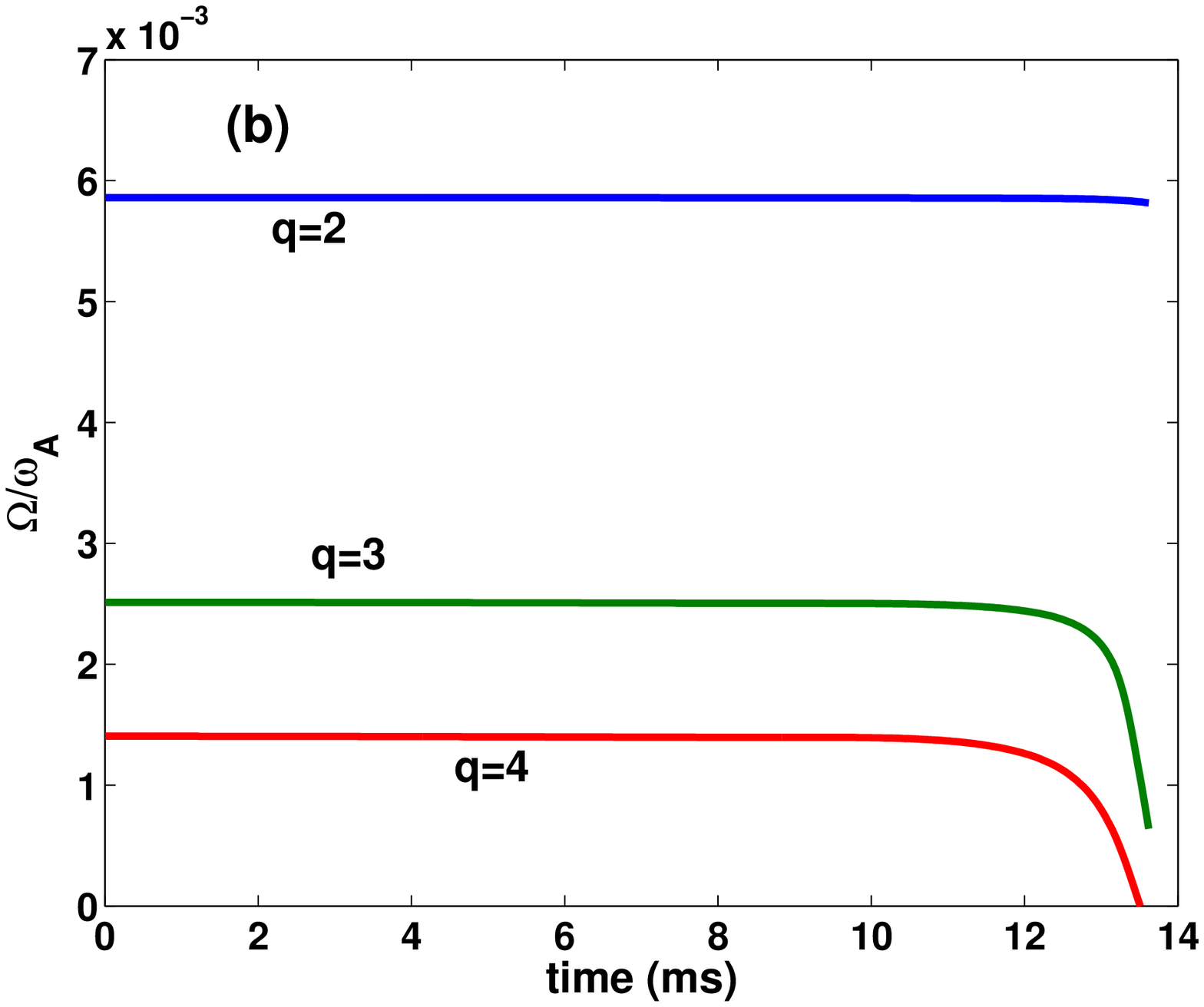}
\caption{Simulated time traces of (a) the net toroidal
  electromagnetic and NTV torques (with reversed sign) acting on the plasma column, and (b)
  the toroidal rotation frequencies
  at the $q=2,3,4$ rational surfaces, for the
  base case.} 
\label{fig:basetime}
\end{center}
\end{figure}

The time traces of the net (integrated over the plasma minor radius)
electromagnetic and NTV torques are compared in
Fig. \ref{fig:basetime}, together with the time traces of the toroidal
rotation frequencies at rational surfaces, for the base case. The net
NTV torque is larger than the net ${\bf j}\times{\bf b}$ torque. But 
during the first $\sim$10ms of the time interval, the amplitudes of
both torques are too small to cause appreciable damping of the flow
(Fig. \ref{fig:basetime}(b)). After about 10ms of simulation, the
amplitudes of both torques rapidly increase, and the toroidal rotation
quickly slows down in the region between the $q=3$ rational surface and
the plasma edge. The full time of the rotational damping (and hence
the RMP penetration) is about 14ms for the base case.           

\subsection{Verification of time stepping scheme}
\begin{figure}
\begin{center}
\includegraphics[width=8cm]{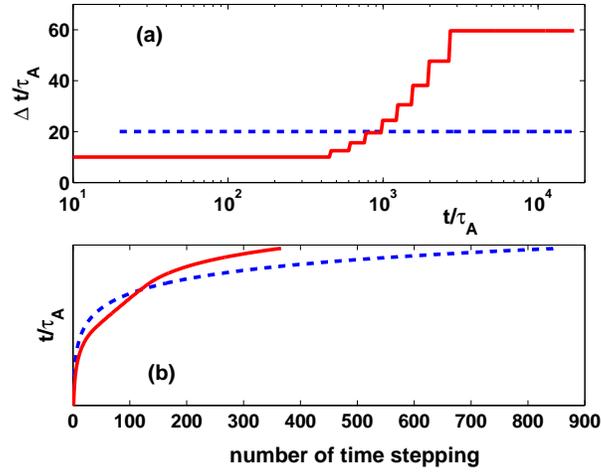}
\caption{Comparison of the simulation history between the adaptive (solid lines)
  and fixed (dashed) time stepping schemes, for the base case: (a) the
  time step $\Delta t$ versus the total simulation time $t$; (b) the total simulation
  time $t$ versus the number of time stepping.} 
\label{fig:cmpts}
\end{center}
\end{figure}

\begin{figure}
\begin{center}
\includegraphics[width=6.5cm]{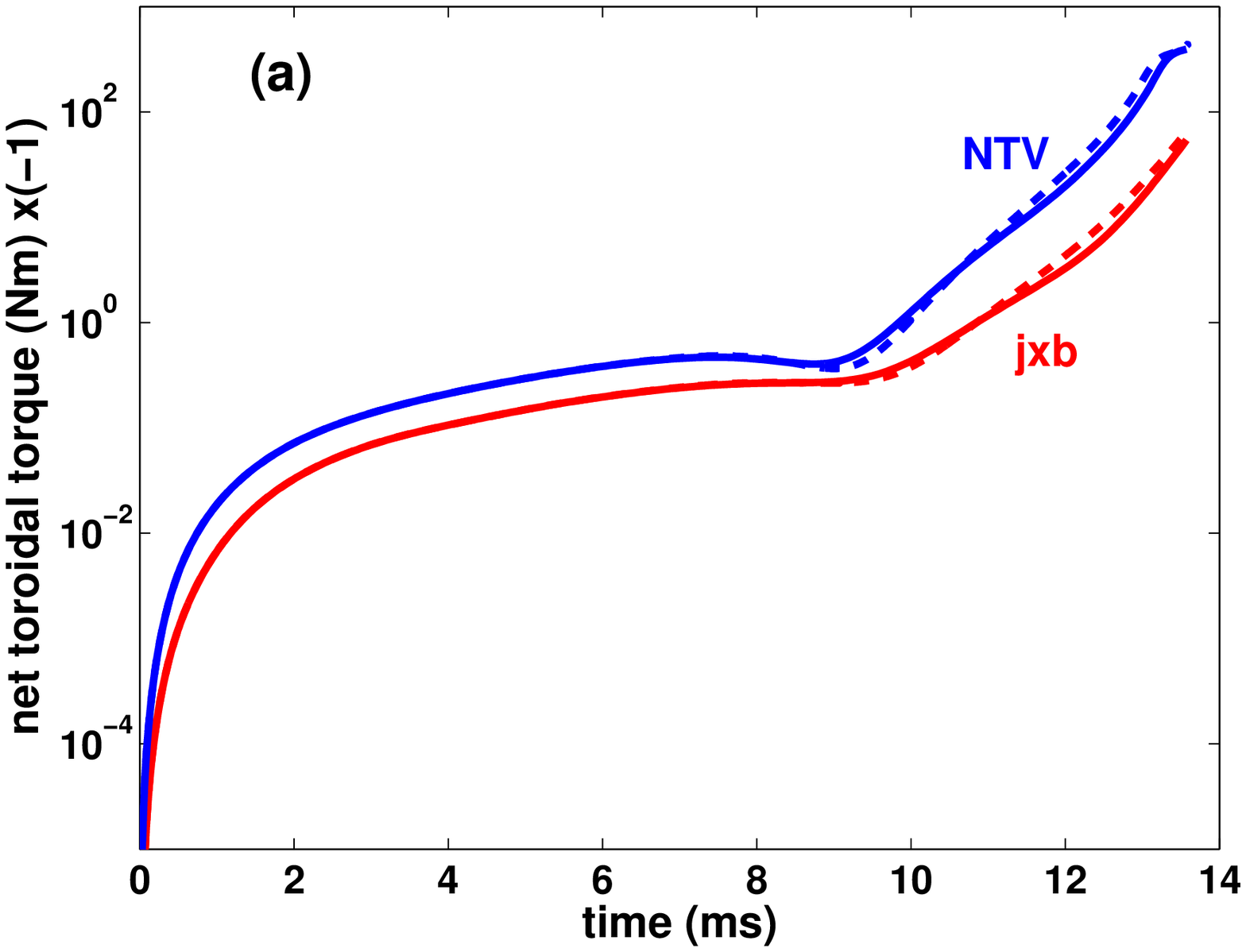}
\includegraphics[width=6.5cm]{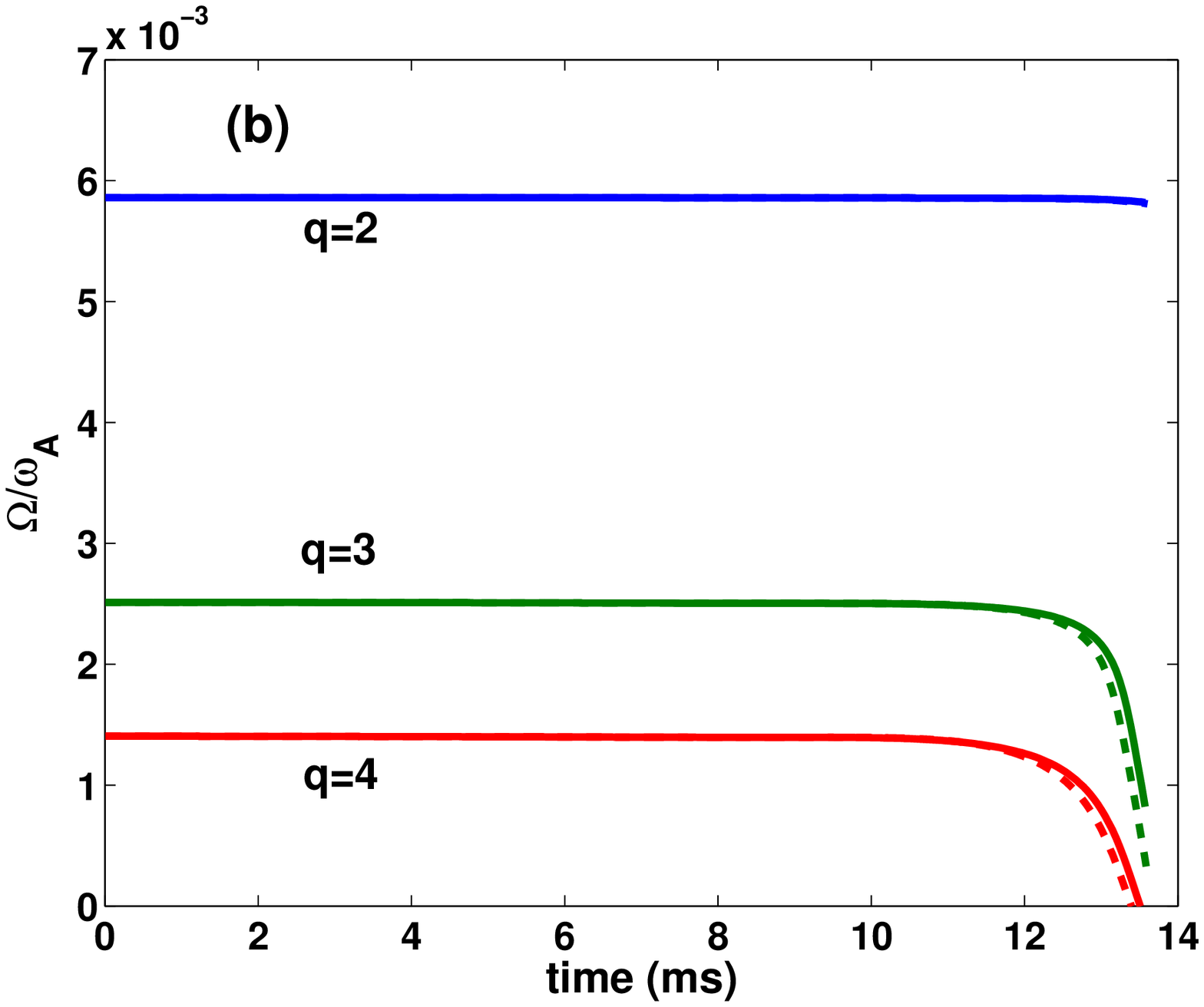}
\caption{Simulated time traces of (a) the net toroidal
  electromagnetic and NTV torques (with reversed sign) acting on the plasma column, and (b)
  the toroidal rotation frequencies
  at the $q=2,3,4$ rational surfaces, for the
  base case with adaptive (solid lines) and fixed (dashed lines) time
  stepping schemes.} 
\label{fig:cmptime}
\end{center}
\end{figure}

For numerical efficiency, we wish to choose as large a time step as possible.
Obviously, the time step cannot be chosen too large, in order
not to affect the field penetration dynamics. A good criterion is that
different choices of the time step should result in the same time
evolution of the numerical solution. For the base case, we use an
adaptive time stepping strategy as described in Section
\ref{sec:adap}. The initial time step (at $t=0$) is set to be
10$\tau_A$. The
time stepping history is shown in Fig. \ref{fig:cmpts} as solid
lines. For this case, the length of the time step steadily increases
during the non-linear evolution. There are also cases where the
length of the time step varies non-monotonically. For comparison, we
run the same case, but with a fixed time 
step of 20$\tau_A$ (dashed lines). The adaptive time stepping scheme
requires much less number of steps to reach the same total simulation
time. More importantly, the numerical solutions, as functions of time,
agree well between two time stepping schemes, as shown in
Fig. \ref{fig:cmptime}.  This demonstrates the validity of our
adaptive scheme.  

\subsection{Numerical results from parametric studies}
\begin{figure}
\begin{center}
\includegraphics[width=6.5cm]{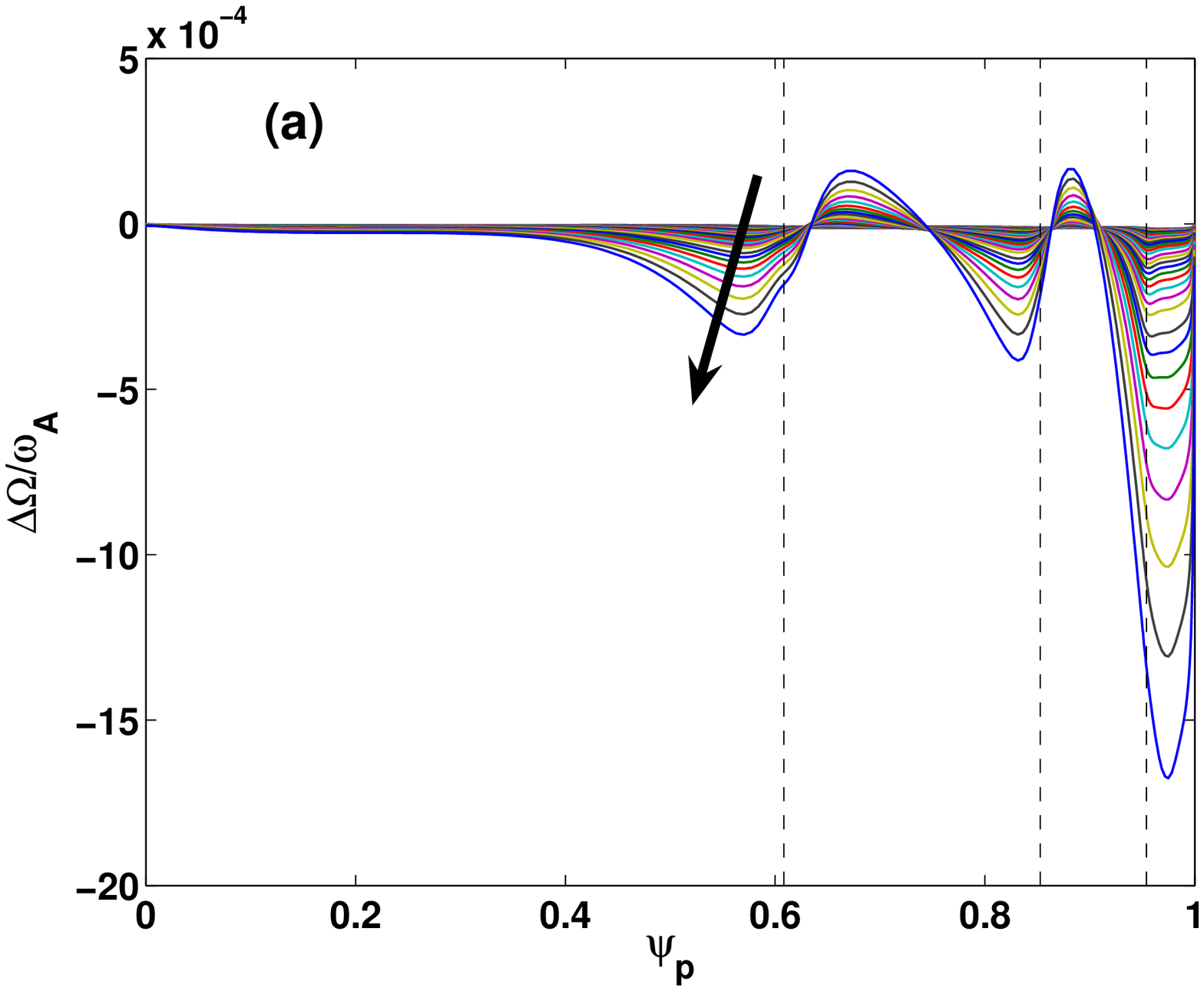}
\includegraphics[width=6.5cm]{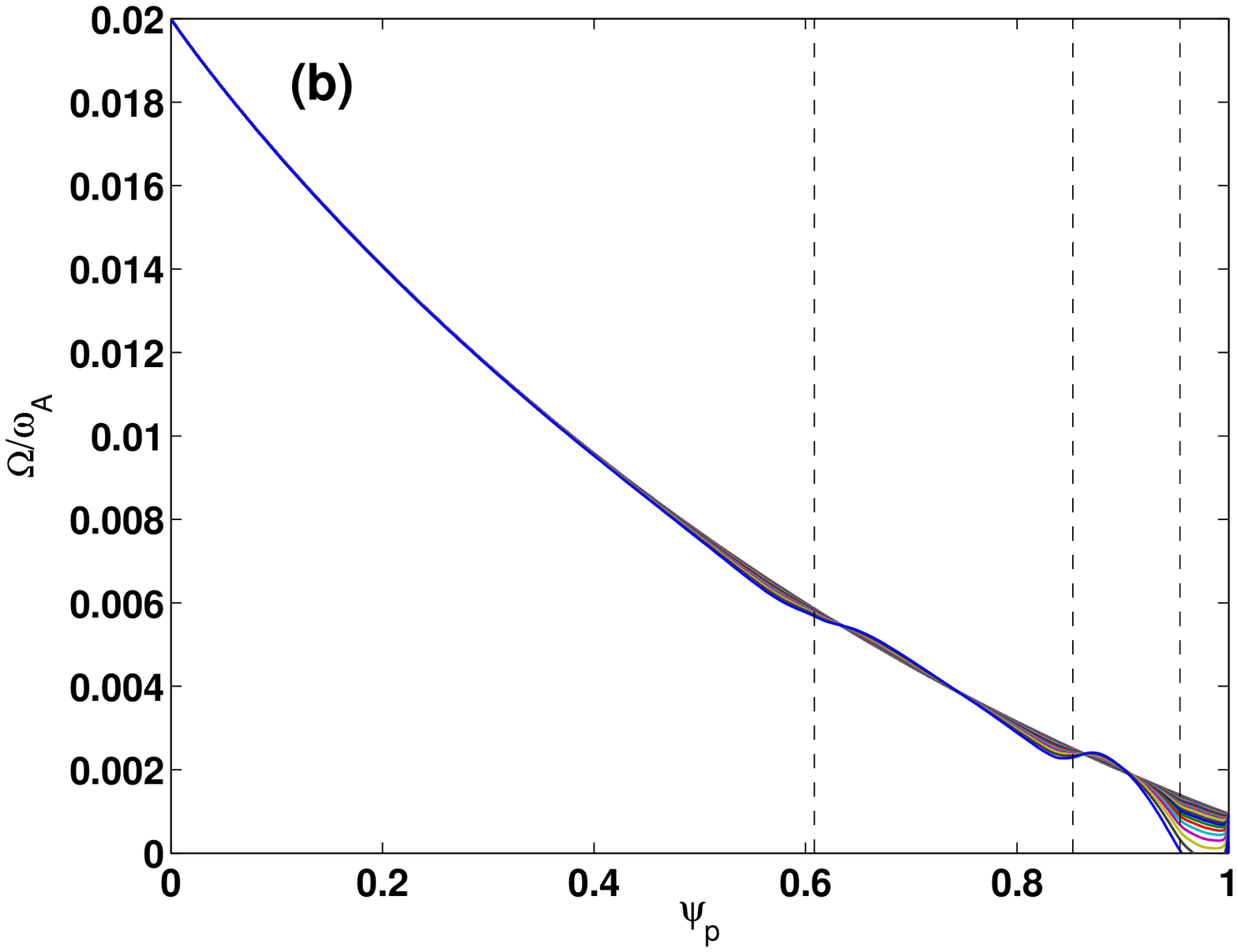}
\caption{Evolution of the simulated radial profiles of (a)
  $\Delta\Omega(\psi_p,t)\equiv\Omega(\psi_p,t)-\Omega(\psi_p,t=0)$ and (b) 
  $\Omega(\psi_p,t)$ for the case without the NTV torque, where $\Omega$ is the toroidal 
  rotation frequency, $\psi_p$ is the normalized equilibrium poloidal
  flux, and $t$ is the time.  Shown are only profiles with a
  time span of 0.1ms, and after 10ms of simulation. The arrow
  indicates the time flow. The vertical dashed lines indicate radial
  locations of the $q=2,3,4$ rational surfaces, respectively.} 
\label{fig:emwp}
\end{center}
\end{figure}

\begin{figure}
\begin{center}
\includegraphics[width=6.5cm]{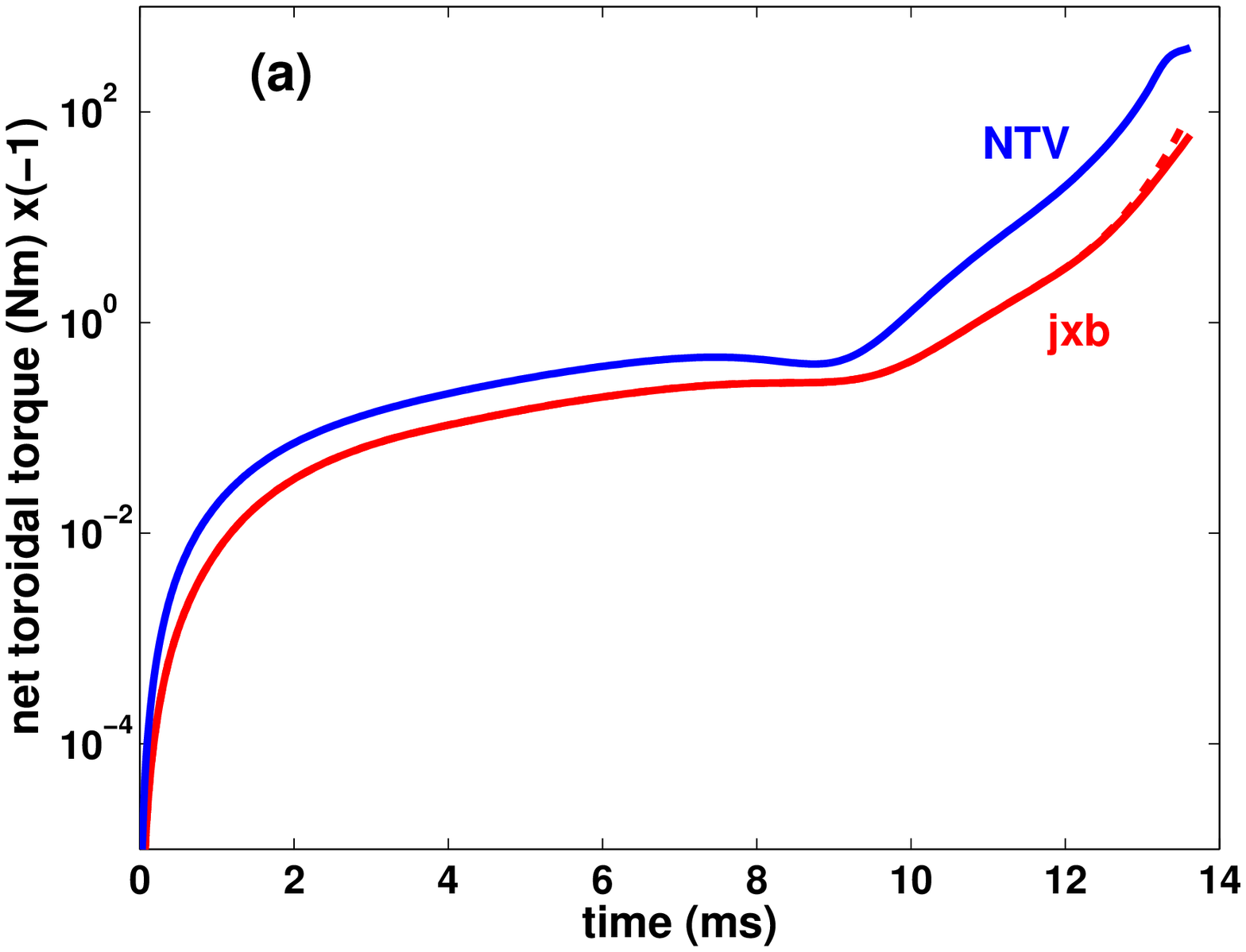}
\includegraphics[width=6.5cm]{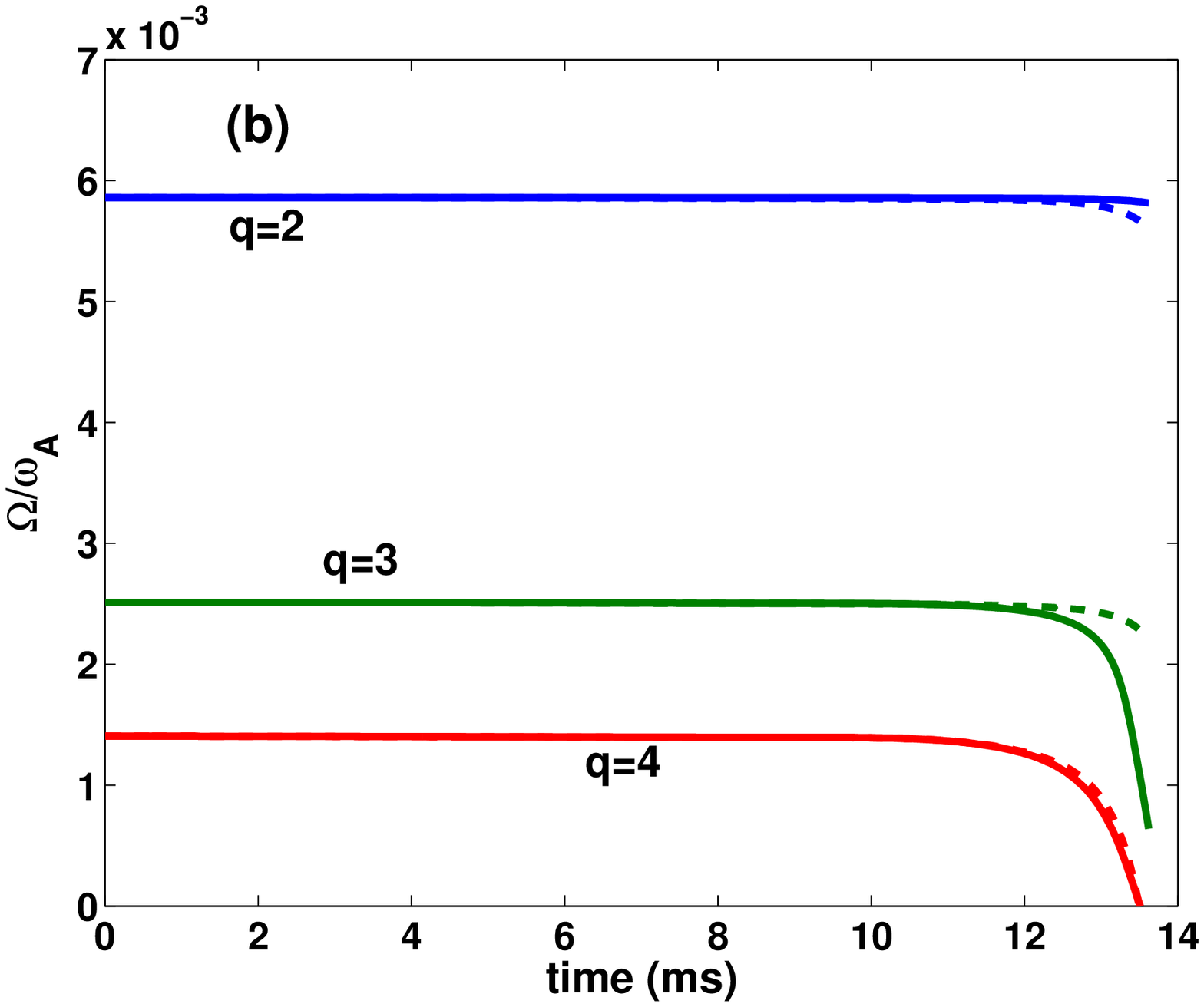}
\caption{Simulated time traces of (a) the net toroidal
  electromagnetic and NTV torques (with reversed sign) acting on the plasma column, and (b)
  the toroidal rotation frequencies
  at the $q=2,3,4$ rational surfaces, for the
  base case (solid lines) and the case without the NTV torque (dashed lines).} 
\label{fig:emtime}
\end{center}
\end{figure}

Figure \ref{fig:basetime} shows that the NTV torque is generally
the dominant momentum sink due to the interaction between the plasma
response with the RMP field, for our plasma and coil configurations. It
is therefore interesting to consider a case without inclusion of the
NTV torque. The results are shown in Figs. \ref{fig:emwp}
and \ref{fig:emtime}, where only the electromagnetic torque is
included in the toroidal momentum balance equation as the sink
term. Compared to the base case, the only significant difference is
that the flow velocity is much less damped between the $q=3$ and 4
rational surfaces in the absence of the NTV torque. As a results, the
full rotation braking (and hence the RMP penetration) occurs near the
very edge of the plasma, mainly outside the $q=4$ rational surface. In
particular, the rotation velocity is still fully damped at the $q=4$
surface, by the ${\bf j}\times{\bf b}$ torque alone. However, the full
damping occurs slightly later (see Fig. \ref{fig:emtime}(b)) than the
base case, where both the 
electromagnetic and the NTV torques have been included into the
momentum equation.  

\begin{figure}
\begin{center}
\includegraphics[width=6.5cm]{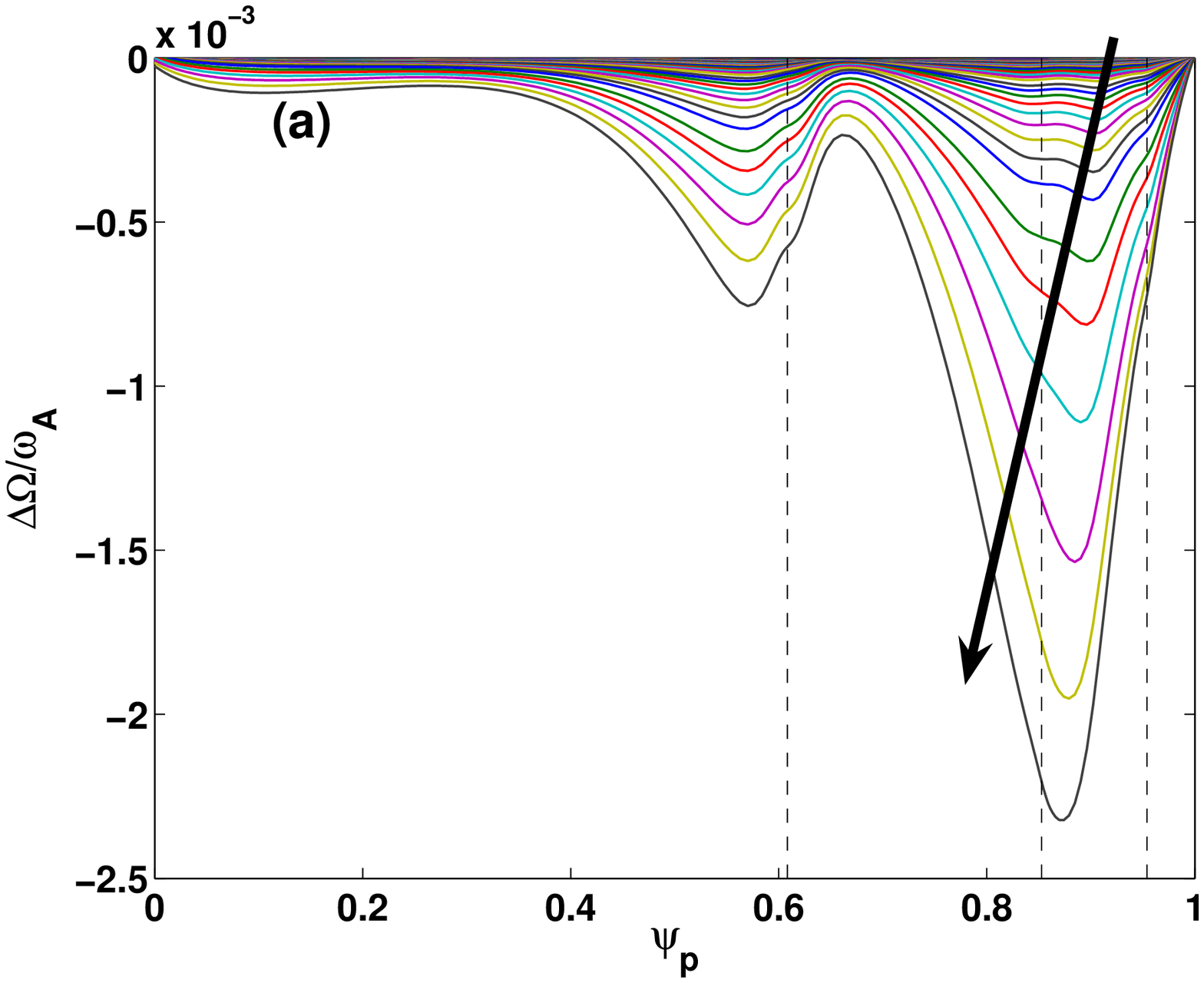}
\includegraphics[width=6.5cm]{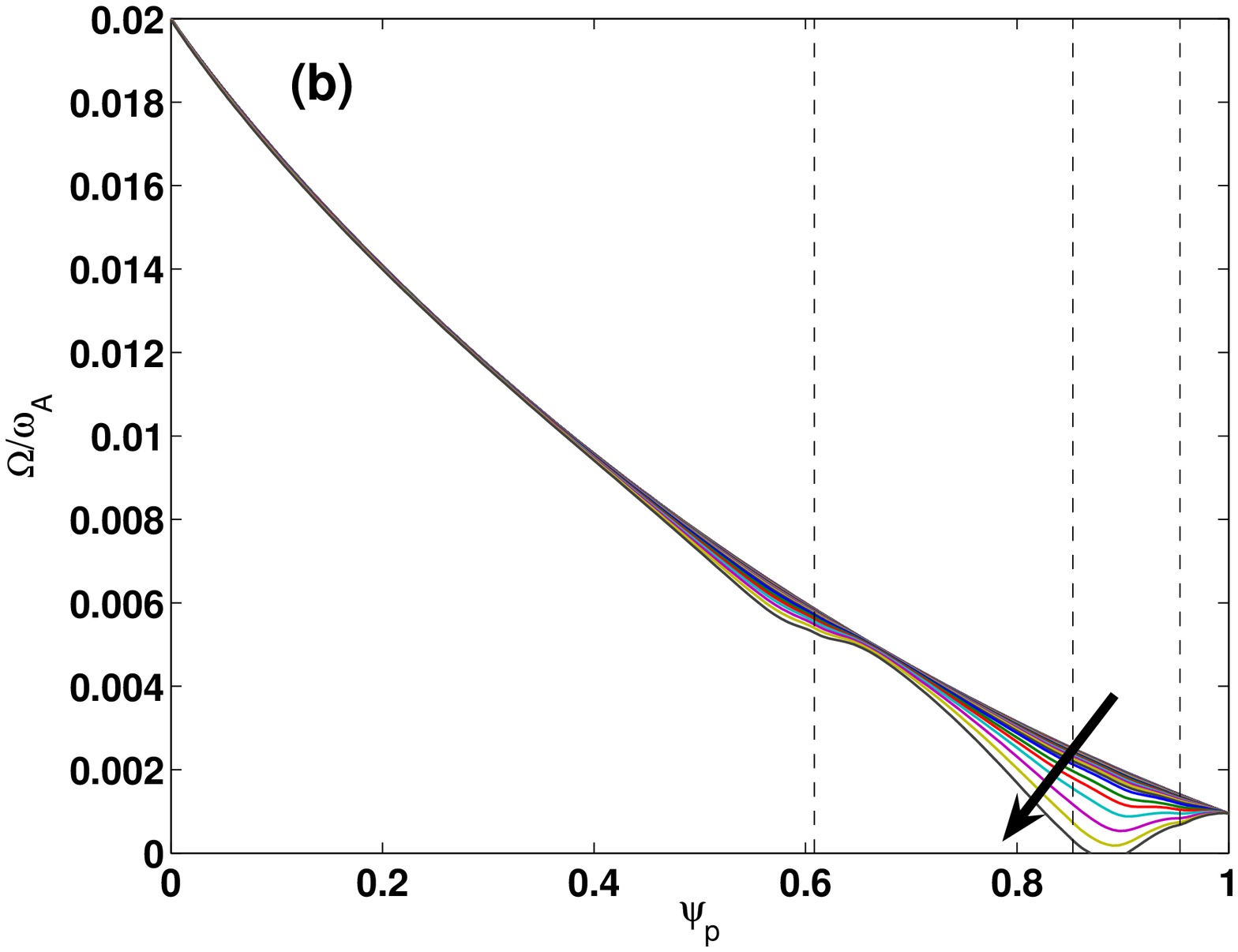}
\caption{Evolution of the simulated radial profiles of (a)
  $\Delta\Omega(\psi_p,t)\equiv\Omega(\psi_p,t)-\Omega(\psi_p,t=0)$ and (b) 
  $\Omega(\psi_p,t)$ for the case with a radially increasing momentum
  diffusion, where $\Omega$ is the toroidal  
  rotation frequency, $\psi_p$ is the normalized equilibrium poloidal
  flux, and $t$ is the time.  Shown are only profiles with a
  time span of 0.1ms, and after 10ms of simulation. The arrow
  indicates the time flow. The vertical dashed lines indicate radial
  locations of the $q=2,3,4$ rational surfaces, respectively.} 
\label{fig:vd3wp}
\end{center}
\end{figure}

\begin{figure}
\begin{center}
\includegraphics[width=6.5cm]{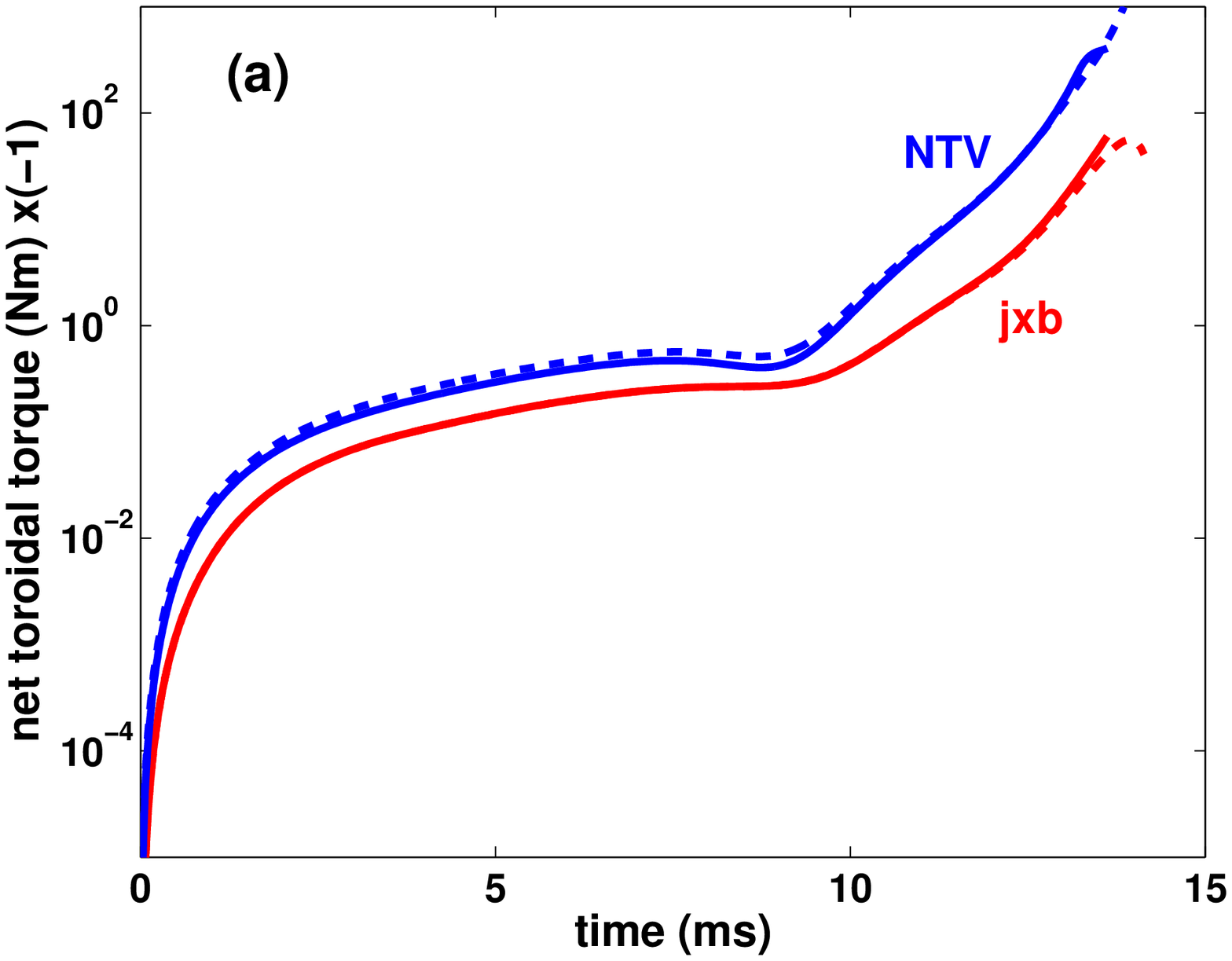}
\includegraphics[width=6.5cm]{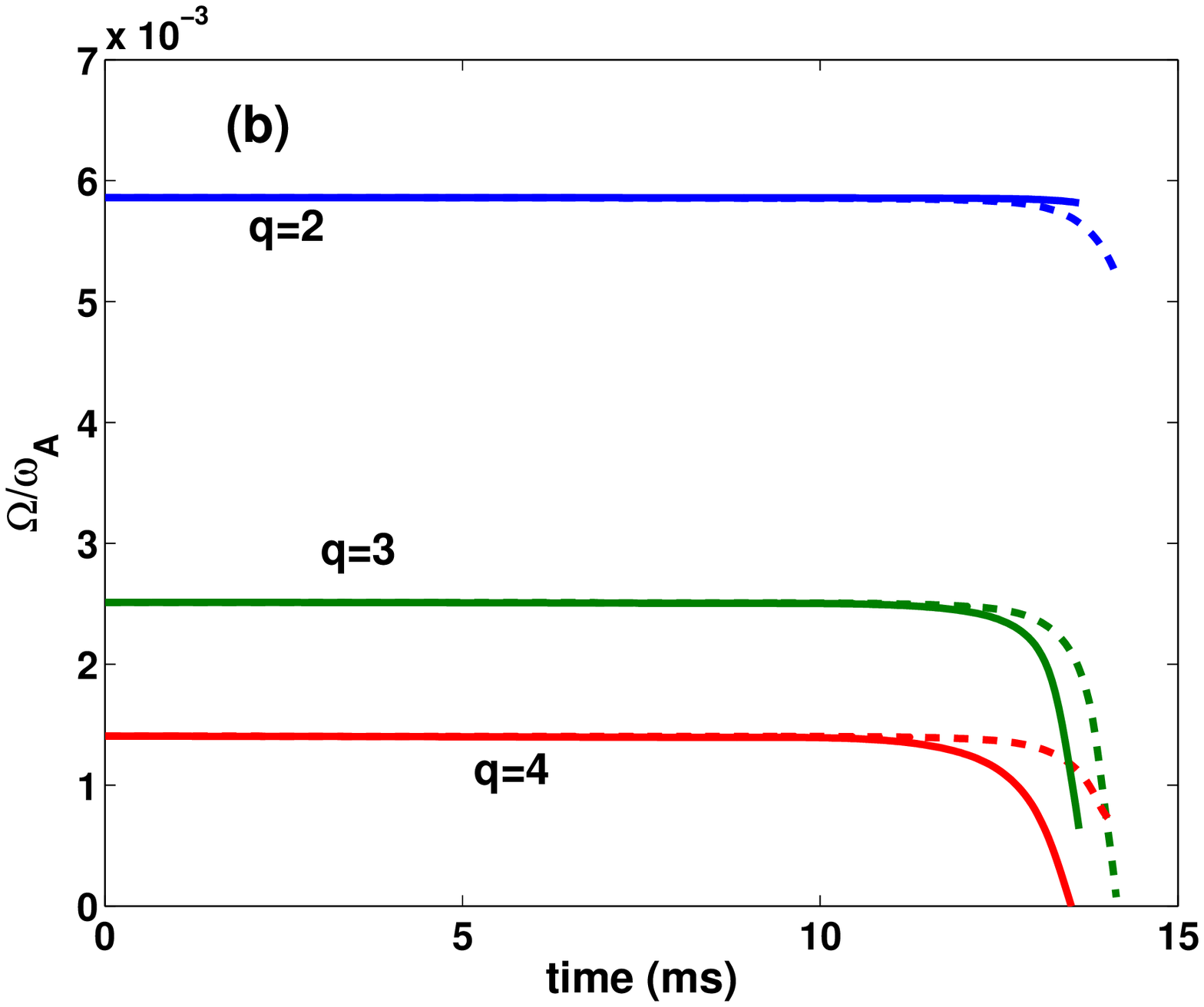}
\caption{Simulated time traces of (a) the net toroidal
  electromagnetic and NTV torques (with reversed sign) acting on the plasma column, and (b)
  the toroidal rotation frequencies
  at the $q=2,3,4$ rational surfaces, for the
  base case (solid lines) and the case with a radially increasing momentum
  diffusion (dashed lines).} 
\label{fig:vd3time}
\end{center}
\end{figure}

The plasma rotation braking, observed in this work, is not very sensitive
to the radial profile of the toroidal momentum diffusion coefficient
$\chi_M(\psi_p)$. In the simulation presented by Figs. \ref{fig:vd3wp}
and \ref{fig:vd3time}, we chose a completely different radial profile
for $\chi_M$, $\chi_M(\psi_p)=\chi_M^0[T_e(\psi_p)/T_e(0)]^{-3/2}$, compared to
the base case, yet the non-linear solutions do not significantly
differ, apart from two observations. (i) Less flow damping is obtained
near the plasma edge as shown in Fig. \ref{fig:vd3wp}(b). This is
because a large momentum diffusion near 
the plasma edge leads to a stronger coupling of the rotation velocity
to the edge boundary condition, which is chosen to be fixed at a small
but finite value. (ii) At all rational surfaces, the rotational
braking occurs slower than the 
base case, as shown in Fig. \ref{fig:vd3time}. We note that the plasma
core rotation is hardly affected by the RMP field, with both (extreme)
types of the toroidal momentum diffusion profiles.    

\begin{figure}
\begin{center}
\includegraphics[width=6.5cm]{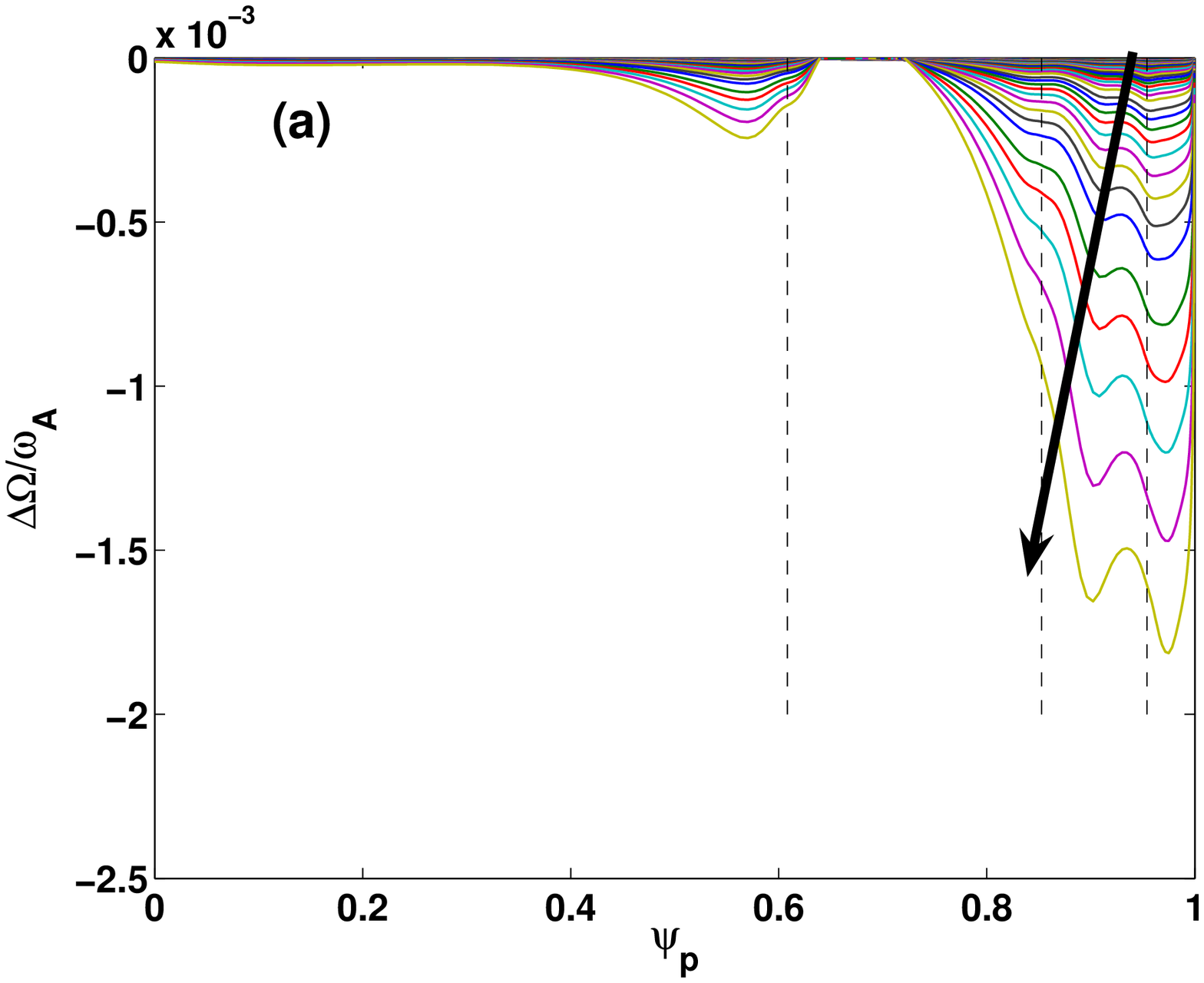}
\includegraphics[width=6.5cm]{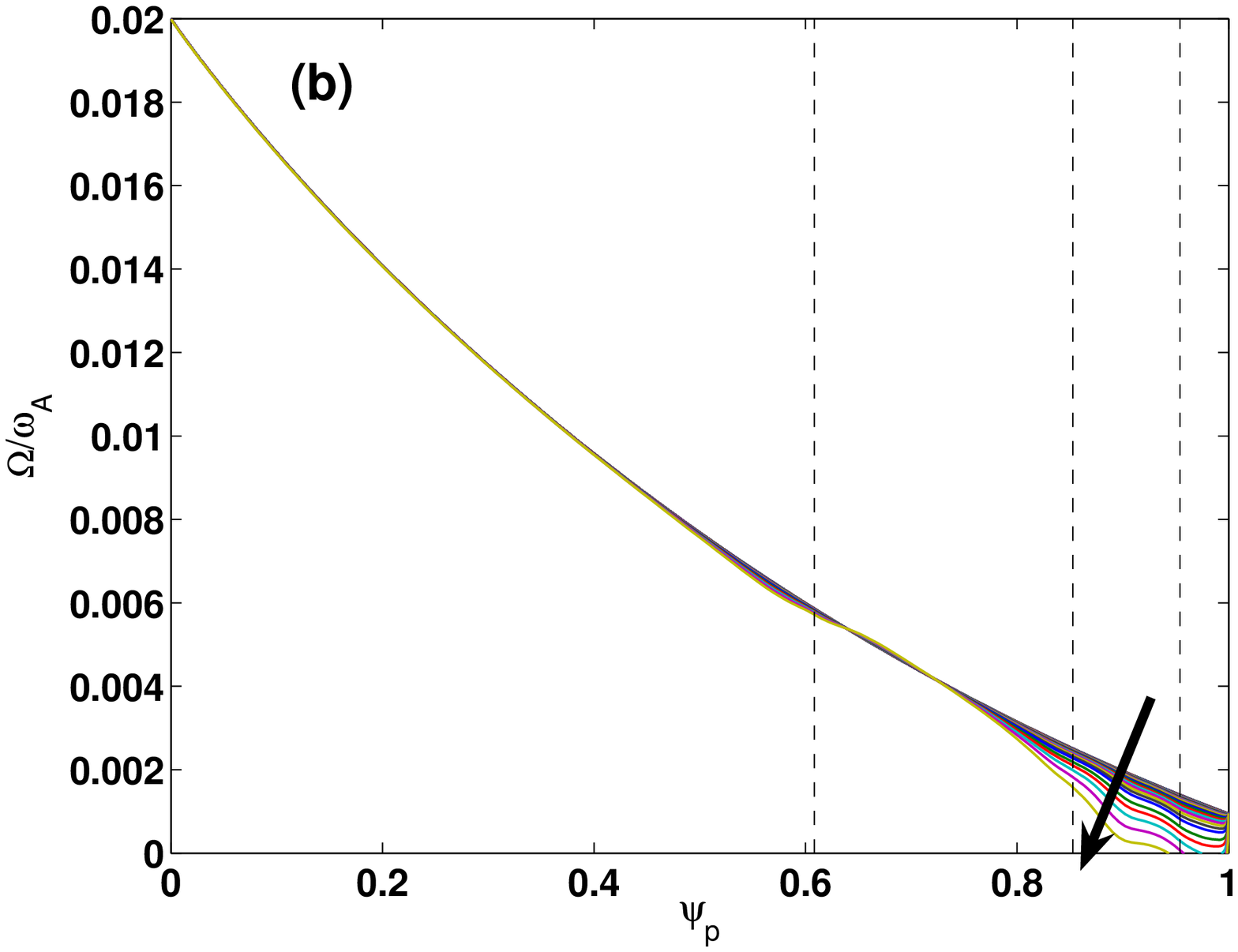}
\caption{Evolution of the simulated radial profiles of (a)
  $\Delta\Omega(\psi_p,t)\equiv\Omega(\psi_p,t)-\Omega(\psi_p,t=0)$ and (b) 
  $\Omega(\psi_p,t)$ for the case with 10kAt coil current, where $\Omega$ is the toroidal 
  rotation frequency, $\psi_p$ is the normalized equilibrium poloidal
  flux, and $t$ is the time.  Shown are only profiles with a
  time span of 0.1ms, and after 10ms of simulation. The arrow
  indicates the time flow. The vertical dashed lines indicate radial
  locations of the $q=2,3,4$ rational surfaces, respectively.} 
\label{fig:fi05wp}
\end{center}
\end{figure}

\begin{figure}
\begin{center}
\includegraphics[width=6.5cm]{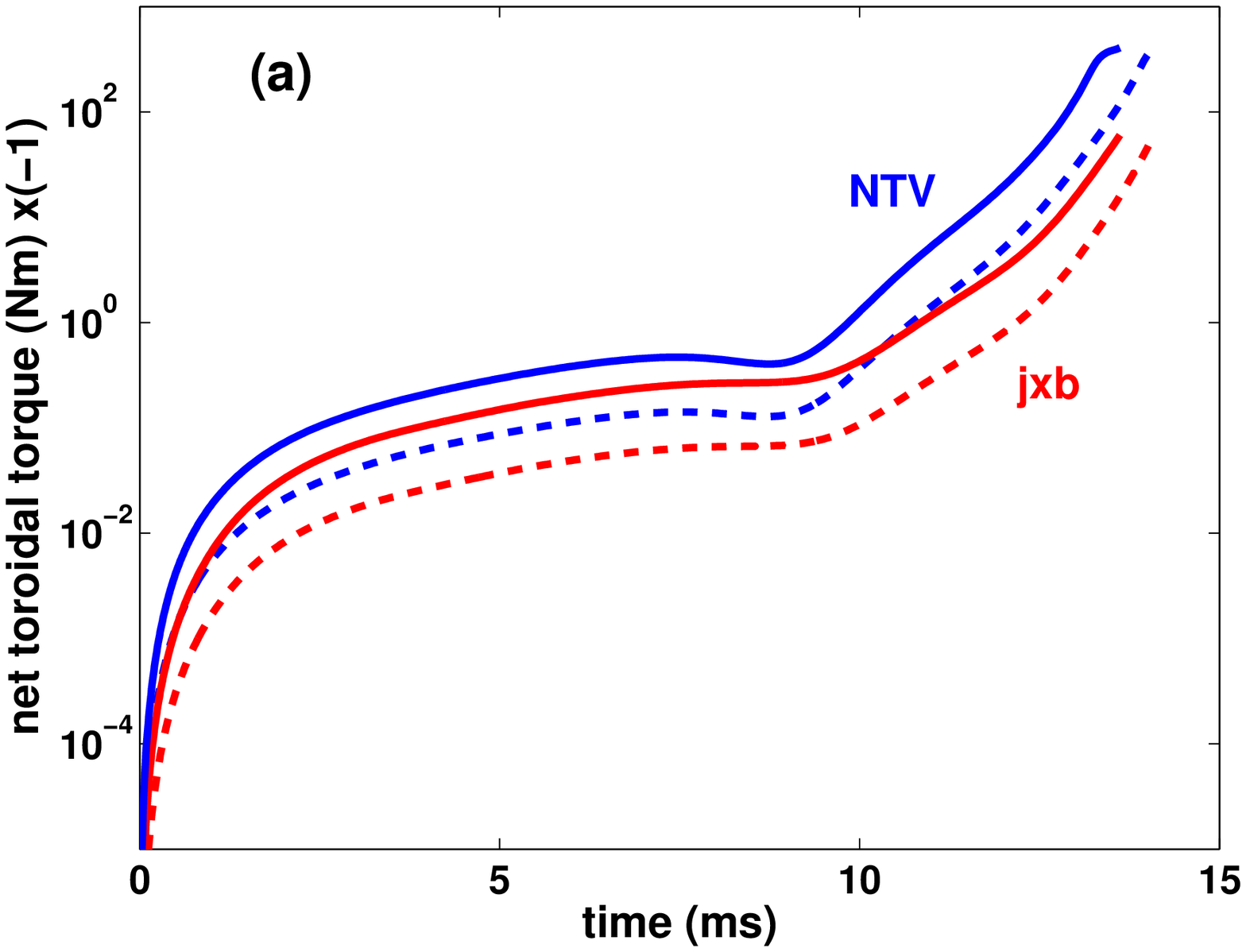}
\includegraphics[width=6.5cm]{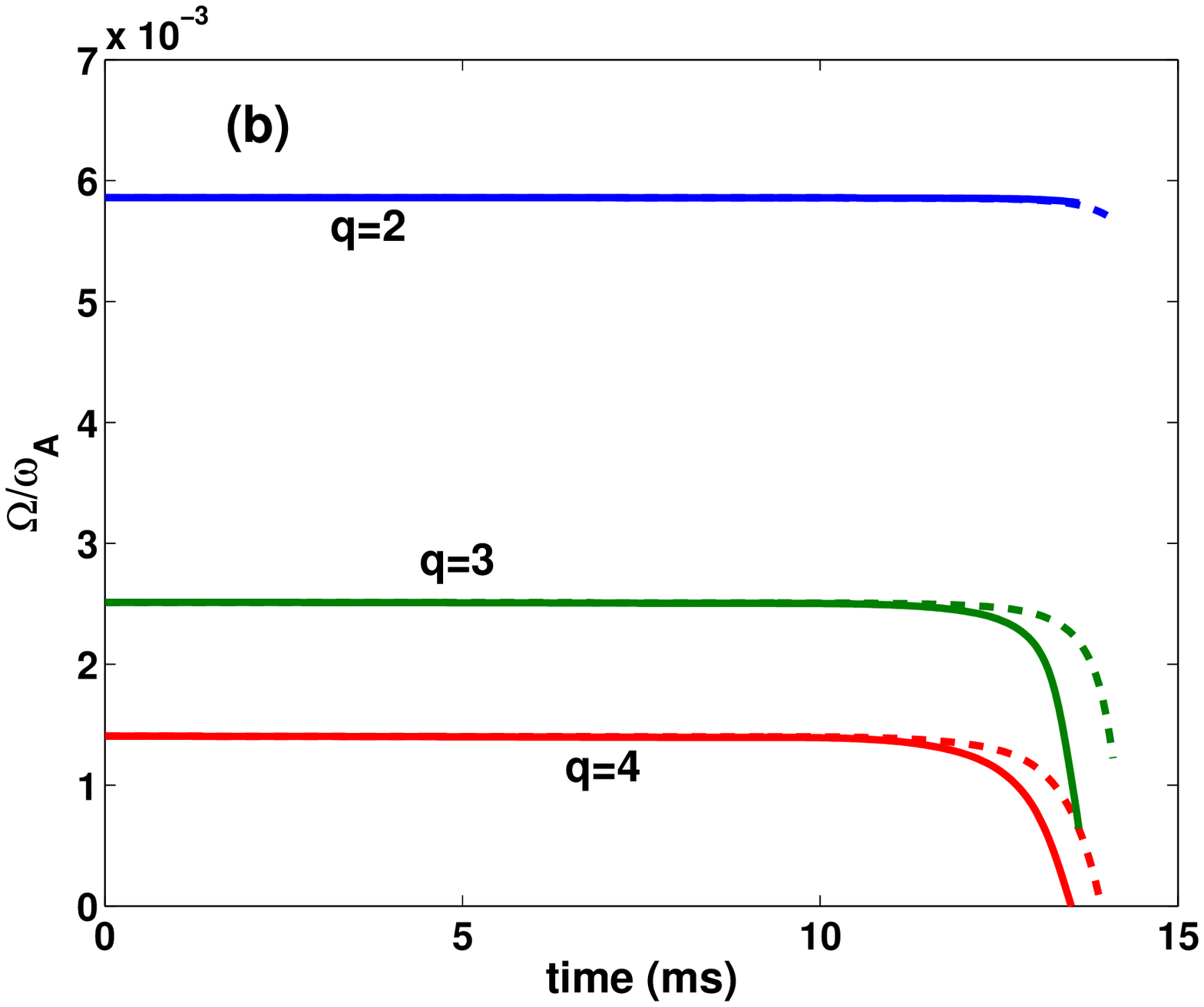}
\caption{Simulated time traces of (a) the net toroidal
  electromagnetic and NTV torques (with reversed sign) acting on the plasma column, and (b)
  the toroidal rotation frequencies
  at the $q=2,3,4$ rational surfaces, for the
  base case (20kAt, solid lines) and the case  with half of the coil
  current (10kAt, dashed lines).} 
\label{fig:fi05time}
\end{center}
\end{figure}

Finally, we also varied the amplitude of the RMP coil current. For
this plasma equilibrium, it appears that even a small amount of
the $n=1$ RMP field can eventually brake the toroidal flow near the
plasma edge. This may be due to the fact that a very low $n$ field is
applied to the plasma. Generally though, as expected, a lower current amplitude
leads to weaker electromagnetic and NTV torques, and to a later
braking of the rotation.  One such example is shown in
Figs. \ref{fig:fi05wp} and \ref{fig:fi05time}, where only half of the
RMP current (i.e. 10kAt) is applied to the plasma, and the simulation
results are compared with the 20kAt case (the base case).

\section{Summary and discussion}
A quasi-linear model is developed to study the RMP field
penetration and the rotation braking in full toroidal geometry. The key physics,
captured by this model, is the non-linear interplay between the
damping of the plasma toroidal rotation by an external RMP field, and the
screening of the RMP field due to the plasma rotation, as a result of
the plasma response to the RMP field. Two toroidal torques - the
electromagnetic ${\bf j}\times {\bf b}$ torque (fluid effect), and the
NTV torque (kinetic effect) - are included in the toroidal momentum
balance equation. An adaptive time stepping scheme is envisaged to
speed up the non-linear simulations, which involves a fully implicit
procedure for solving the MHD equations. 

For a test toroidal equilibrium with H-mode plasma, we find that a
$n=1$ RMP field does not significantly change the plasma core
rotation, {\it before} fully braking the rotation near the plasma edge
region, most often outside the $q=3$ rational surface. This
observation does not exclude the core rotation damping in a longer
time scale. However, our (thin island) model breaks down after the
full damping the edge flow. 
  
The toroidal computations quantify several factors affecting the
dynamics of the RMP field penetration. (i) The
plasma response to RMP fields induces a larger net NTV torque, than
the ${\bf
  j}\times {\bf b}$ torque. This is not a ubiquitous observation, but
does occur for the equilibrium considered in this work. Moreover, the
NTV torque provides predominant flow damping between the $q=3$ and 4
rational surfaces.   (ii) Not surprisingly, we
find that a larger RMP 
amplitude leads to stronger rotational damping and faster field
penetration. The penetration time is generally in the order of ten
milliseconds for our example.  (iii) The radial profile of the
momentum diffusion coefficient, which is an uncertain factor in our
simulations, does not play a significant role for the flow damping
observed in this study. 

{For the cases considered in this work, no steady state solution
is found, although steady solutions are found by MARS-Q for other
plasmas \cite{LiuPPCF12}. The boundary condition, assumed for the
momentum balance equation at the plasma boundary, also affects the
achievement of the steady state solution. For instance, by assuming a
Neumann type of boundary condition,
MARS-Q simulation can lead to steady state solutions. But these
solutions are physically less relevant.}

{Even though the results presented in the paper mainly demonstrate
the rotational braking effect due to the applied RMP field, it is
worthwhile to further discuss some key aspects of the RMP field
penetration itself, in particularly the penetration mechanism. In our
model, the field penetration process is dictated by the strong
non-linear interplay between the resistive plasma response and the
toroidal flow damping. Therefore, the penetration time is eventually
associated, from one side, with the resistive decay of the current sheets, formed near
rational surfaces that tend to prevent the penetration of resonant
field components, and from the other side, with the diffusion of the
toroidal momentum. The scaling of the penetration time versus basic
plasma and coil parameters, which has not been established in this
initial work but will be systematically investigated in the future, is
associated with these physics. For instance, we mention that a linear
scaling of the penetration time, versus the magnetic Lundquist number,
has been established in a cylindrical simulation \cite{Becoulet09}. No
scaling has been established with respect to the plasma initial flow
speed, though a qualitative understanding is possible relying on the
following two arguments: (i) a slower initial flow (before applying
the RMP field) normally yields less screening of the resonant field
perturbations, and hence should facilitate the field penetration; (ii)
at sufficiently slow rotation, the ${\bf E}\times{\bf B}$ flow
frequency can be in resonance with the precessional drift frequency of
trapped thermal particles, resulting in enhanced (resonant) NTV
torque, which in turn can lead to a faster damping of the flow and
hence the field penetration.}

{Another interesting question is whether the penetration time is
associated with the Alfv\'en time, expected for establishing a magnetic
equilibrium. It appears that both experimental evidence
\cite{LiuPPCF12} and the numerical results shown in this work, as well
as other theoretical work} \cite{Becoulet09,Park10}, {indicate that the
resonant component of the applied magnetic 
field penetrates into the plasma in the milliseconds time scale, much slower than the
Alfv\'en time.}  
   
We point out that the present study is based on a single
fluid plasma model. It can be argued that the electron response may be
important in the RMP field shielding. Therefore, a two-fluid model, or
even a full kinetic model \cite{Park10}, may be necessary to better describe the
plasma behavior in the presence of RMP fields. The
possible field line stochastisation can induce an additional plasma
radial current \cite{Rozhansky10}, and consequently field
screening. These effects have not been taken into account in our present
quasi-linear model.

{\bf Acknowledgments.}
YQL thanks Drs. I.T. Chapman, B.D. Dudson, G. Fishpool, R.J. Hastie, T.C. Hender,
D.F. Howell, E. Nardon, V.D. Pustovitov, S. Saarelma, and A.J. Webster for very
helpful discussions during this work, in particular Dr. Fishpool for
suggesting the Dirichlet boundary condition for the momentum solver,
and Dr. Chapman for many helpful suggestions improving the manuscript.

This work was part-funded by the RCUK Energy Programme under grant
EP/I501045 and the European Communities under the contract of
Association between EURATOM and CCFE. The views and opinions expressed
herein do not necessarily reflect those of the European Commission.

{Youwen Sun would like to acknowledge the support from the National
Magnetic Confinement Fusion Science Program of China under Grant
No. 2013GB102000 and No. 2012GB105000, and the National Natural
Science Foundation of China under Grant No. 11205199 and No. 10725523.}  

{We also thank the anonymous reviewer for interesting comments, that lead
to discussions of the important issues associated with the field
penetration mechnism.}


\end{document}